%% file: zzz-main-sigconf.tex
\documentclass[sigconf,screen]{acmart}

\usepackage{xcolor} 
\usepackage[nomessages]{fp} 
\usepackage{booktabs} 
\usepackage{multirow} 
\usepackage{url}
\usepackage{framed}
\usepackage{enumitem}
\usepackage{array}
\usepackage{tabularx}
\usepackage{balance}
\usepackage{amsmath} 
\usepackage{siunitx} 
\usepackage{subcaption}
\usepackage{adjustbox}
\usepackage[flushleft]{threeparttable}
\usepackage{setspace} 
\usepackage{caption} 

\definecolor{ForestGreen}{rgb}{0.13, 0.55, 0.13}
\definecolor{tomBlue}{HTML}{0196CE}

\usepackage{soul,color}
\soulregister\cite7
\soulregister\ref7
\soulregister\pageref7
\setstcolor{red}
\setul{}{.2ex}

\newcolumntype{L}[1]{>{\raggedright\let\newline\\\arraybackslash\hspace{0pt}}m{#1}}
\newcolumntype{C}[1]{>{\centering\let\newline\\\arraybackslash\hspace{0pt}}m{#1}}
\newcolumntype{R}[1]{>{\raggedleft\let\newline\\\arraybackslash\hspace{0pt}}m{#1}}

\usepackage{xspace}
\newcommand{\three}{S$_{3}$\xspace}
\newcommand{\six}{S$_{6}$\xspace}
\newcommand{\onehundred}{S$_{100}$\xspace}

\newcommand{\politifact}{\texttt{PolitiFact}\xspace}
\newcommand{\abc}{\texttt{ABC}\xspace}

\newcommand{\politifactzero}{\texttt{pants-on-fire}\xspace}
\newcommand{\politifactlie}{\texttt{lie}\xspace}
\newcommand{\politifactone}{\texttt{false}\xspace}
\newcommand{\politifacttwo}{\texttt{barely-true}\xspace}
\newcommand{\politifactthree}{\texttt{half-true}\xspace}
\newcommand{\politifactfour}{\texttt{mostly-true}\xspace}
\newcommand{\politifactfive}{\texttt{true}\xspace}

\newcommand{\abczero}{\texttt{negative}\xspace}
\newcommand{\abcone}{\texttt{in-between}\xspace}
\newcommand{\abctwo}{\texttt{positive}\xspace}

\newcommand{\high}{\texttt{H}\xspace}
\newcommand{\low}{\texttt{L}\xspace}

\newcommand{\myparagraph}[1]{\vspace{0.5\baselineskip}\noindent{\textit{#1}}~}

\AtBeginDocument{%
  \providecommand\BibTeX{{%
    \normalfont B\kern-0.5em{\scshape i\kern-0.25em b}\kern-0.8em\TeX}}}

\copyrightyear{2020}
\acmYear{2020}
\setcopyright{acmlicensed}\acmConference[SIGIR '20]{Proceedings of the 43rd
International ACM SIGIR Conference on Research and Development in
Information Retrieval}{July 25--30, 2020}{Virtual Event, China}
\acmBooktitle{Proceedings of the 43rd International ACM SIGIR Conference on
Research and Development in Information Retrieval (SIGIR '20), July 25--30,
2020, Virtual Event, China}
\acmPrice{15.00}
\acmDOI{10.1145/3397271.3401112}
\acmISBN{978-1-4503-8016-4/20/07}

\author{Kevin Roitero}
\email{roitero.kevin@spes.uniud.it}
\affiliation{
\institution{University of Udine}
\city{Udine}
\country{Italy}
}

\author{Michael Soprano}
\email{michael.soprano@outlook.com}
\affiliation{
\institution{University of Udine}
\city{Udine}
\country{Italy}
}

\author{Shaoyang Fan}
\email{fsysean@gmail.com}
\affiliation{
\institution{The University of Queensland}
\city{Brisbane}
\country{Australia}
}

\author{Damiano Spina}
\email{damiano.spina@rmit.edu.au}
\affiliation{
\institution{RMIT University}
\city{Melbourne}
\country{Australia}
}

\author{Stefano Mizzaro}
\email{mizzaro@uniud.it}
\affiliation{
\institution{University of Udine}
\city{Udine}
\country{Italy}
}

\author{Gianluca Demartini}
\email{demartini@acm.org}
\affiliation{
\institution{The University of Queensland}
\city{Brisbane}
\country{Australia}
}

\begin{document}
\fancyhead{}


\title[Can The Crowd Identify Misinformation Objectively? The Effects of Judgment Scale and Assessor's Background]{Can The Crowd Identify Misinformation Objectively?\\ The Effects of Judgment Scale and Assessor's Background}

\begin{abstract}
\input{0-abstract.tex}
\end{abstract}





\maketitle

\input{1-intro.tex}

\input{2-relwork.tex}

\input{3-method.tex}
\input{4-Analysis.tex}
\input{5-concl.tex}

\begin{acks}
This work is partially
supported by 
a \grantsponsor{}{Facebook Research}{https://research.fb.com/programs/research-awards/proposals/the-online-safety-benchmark-request-for-proposals/} award
and by an \grantsponsor{DP190102141}{Australian Research Council
Discovery Project (DP190102141)}{http://purl.org/au-research/grants/arc/DP190102141}.
We thank Devi Mallal from RMIT ABC Fact Check for facilitating access to the \abc dataset.
\end{acks}


\newpage
\bibliographystyle{ACM-Reference-Format}
\bibliography{zzz-reference}

\end{document}

%% file: 0-abstract.tex
Truthfulness judgments are a fundamental step in the process of fighting misinformation, as they are crucial to train and evaluate classifiers that automatically distinguish true and false statements. Usually such judgments are made by experts, like journalists for political statements or medical doctors for medical statements. In this paper, we follow a different approach and rely on (non-expert) crowd workers. This of course leads to the following research question: Can crowdsourcing be reliably used to assess the truthfulness of information and to create large-scale labeled collections for information credibility systems? To address this issue, we present the results of an extensive study based on crowdsourcing: we collect thousands of truthfulness assessments over two datasets, and we compare expert judgments with crowd judgments, expressed on scales with various granularity levels. We also measure the political bias and the cognitive background of the workers, and quantify their effect on the reliability of the data provided by the crowd.

%% file: 1-intro.tex
\section{Introduction}
The information we are exposed to influences our decision making processes.
Thus, understanding what information should be trusted and which should not be trusted is key for democracy processes to function as supposed to.
The research community has been focusing on developing techniques and datasets to automatically classify the credibility of information \cite{politifact,conroy2015automatic}. Key approaches to automatically differentiate between false and valid claims also include neural models \cite{wang2018eann,singhania20173han,ruchansky2017csi}.
Information credibility assessment (also known under the name of \textit{fact checking}) is a task that has  gained popularity due to the spread of misinformation online. This is often done with the intended mean of deceiving people towards a certain political agenda.
The task of checking the veracity of published information has been traditionally performed by expert fact checkers, that is, journalists who perform the task by verifying information sources and searching for evidence that supports the claims made by the document or statement they are verifying.
On the other hand, the need for manual fact checking is rapidly growing also due to the increasing volume of misleading and false information online \cite{vo2018rise}.
Because of this, it becomes infeasible for journalists to provide fact checking results for all news which are being continuously  published.
Moreover, relying on fact checking results  requires  trusting those who performed the fact checking job. This is something the average web user may not be willing to accept.
Thus, a more scalable and decentralized approach would allow fact checking to be more widely available.

In this paper, we study the limitations of non-expert fact checkers identifying misinformation online.
We run a very large crowdsourcing experiment where we ask crowd workers to fact check statement given by politicians and search for evidence of statement validity using a custom web search engine that we control.
Previous work has looked at the effect of workers' cognitive bias on the quality of crowdsourced relevance judgments \cite{eickhoff2018cognitive} and subjectivity labels \cite{hube2019understanding}. 
In this work we look at human bias on fact checking tasks where the effect may be even stronger given the opinionated dimension of the analyzed content.
In our experiments we collected data on the assessors' political background and cognitive abilities, and control for the political standing of the statements to be fact checked, the geographical relevance of the statements, the assessment scale granularity, and the truthfulness level.
We use one dataset with statements given by US politicians and one dataset with statements given by Australian politicians and ask US-based crowd workers to perform the fact checking task. For each dataset we also used available expert assessments to compare crowd assessments against. In this way, we are able to observe how each crowd worker bias is reflected in the data they generated as we assume US-based workers might have knowledge of US politics but less likely would have knowledge of Australian politics in terms of political figures and topics of discussion.

We investigate the following Research Questions:

\begin{itemize}
    \item[RQ1:] Are the used assessment scales suitable to gather, by means of crowdsourcing, truthfulness labels on political statements?
    \item[RQ2:] Which is the relationship and the agreement between the crowd and the expert labels? And between the labels collected using different scales?
    \item[RQ3:] Which are the sources of information that crowd workers use to identify online misinformation? 
    \item[RQ4:] Which is the effect and the role of assessors' background in objectively identify online misinformation?
\end{itemize}

Our results show that the scale used to collect the judgments does not affect their quality; the agreement among workers is low, but when properly aggregating workers' answers and merging truthfulness levels crowdsourced data correlates well with expert fact checker assessments. The crowd assessors' 
background has an impact on the judgments they provide.
To the best of our knowledge, the dataset used for this paper (\url{https://github.com/KevinRoitero/crowdsourcingTruthfulness}) is the first dataset containing truthfulness assessments produced by the crowd on multiple scales.

The rest of the paper is structured as follows. 
Section \ref{sec:relwork} presents a summary of 
the relevant recent work.
In Section \ref{sec:exp} we present our study setup discussing the used datasets, the fact checking task design, and the assessment scales we consider.
Section~\ref{sec:res} presents our observations on the quality of the data provided by crowd workers, the agreement between crowd workers and experts, and how workers' bias impact their data.
Finally, in Section \ref{sec:conc} we discuss our key findings and draw our conclusions.

%% file: 2-relwork.tex
\section{Related Work}\label{sec:relwork}

In the last few years, the research community has been looking at automatic check-worthiness predictions \cite{gencheva2017context,vasileva2019takes}, at truthfulness detection/credibility assessments \cite{Popat_2019, mihaylova2019semeval, atanasova2019automatic, clef2018checkthat, elsayed2019overview,kim2019homogeneity}, and at developing fact-checking URL recommender systems and text generation models to mitigate the impact of fake news in social media \cite{vo2018rise, vo2019learning, you2019attributed}. In this section we focus on the literature that explored crowdsourcing methodologies to collect truthfulness judgments, the different judgment scales that have been used so far, and the relation between assessors' bias and the data they produce.
    
\myparagraph{Crowdsourcing Truthfulness.}
Crowdsourcing has become a popular methodology to collect human judgments and has been used in the context of information credibility research.
For example, 
\citet{Kriplean:2014:IOF:2531602.2531677}  analyzed volunteer crowdsourcing when applied to  fact-checking.
\citet{zubiaga2014tweet} looked at disaster management and asked 
crowd
workers to assess the credibility of tweets. Their results show that it is difficult for crowd workers to properly assess information truthfulness, but also that the source reliability is a good indicator of trustworthy information.
Related to this, the Fact Checking Lab at CLEF~\cite{clef2018checkthat,elsayed2019overview} addressed the task of ranking sentences according to their need to be fact-checked.
\citet{INRA:2018} looked at assessing news quality along eight different quality dimensions using crowdsourcing.
 \citet{RSDM:2018} and 
\citet{labarbera2020crowdsourcing} 
recently studied how users perceive fake news statements.
As compared to previous work that  looked at crowdsourcing  information credibility tasks, we look at the impact of  assessors' background  and rating scales on the quality of the truthfulness judgments they provide.

\myparagraph{Judgment Scales.} 
Fact-checking websites collect a large number of high-quality labels generated by experts.
However, each fact-checking site and dataset defines its own labels and rating system used to describe the authenticity of the content.
Therefore, to integrate multiple datasets, 
converging to a common rating scale becomes very important.
\citet{vlachos2014fact} aligned labels from Channel 4 and PolitiFact to a five-point scale: \texttt{False}, \texttt{Mostly\hspace{0.3em}False}, \texttt{Half\hspace{0.3em}True}, \texttt{Mostly\hspace{0.3em}True}, and \texttt{True}.
\citet{clef2018checkthat}
retrieve evaluations of different articles at \url{factcheck.org} to assess claims made in American political debates.
They then generate  labels on a three-level scale: \texttt{False}, \texttt{Half\hspace{0.3em}True}, and \texttt{True}. 
\citet{vosoughi2018spread} check the consistency between multiple fact-checking websites on three levels: \texttt{False}, \texttt{Mixed}, and \texttt{True}.
\citet{tchechmedjiev2019claimskg} 
look at rating distributions over different datasets and define a standardized scoring scheme using four 
categories: \texttt{False}, \texttt{Mixed}, \texttt{True}, and \texttt{Other}. 
For these works, we can conclude that different datasets have been using different scales and that meta-analyses have tried to merge scales and aggregate ratings together. While no clear preferred scale has yet emerged, there seems to be a preference towards coarse-grained scales with just a few (e.g., three or four) levels as they may be more user-friendly when labels need to be interpreted by information consumers.
In our work we use the original dataset scale at six levels and compare ratings collected with that scale against a more coarse-grained scale (i.e., three levels) and a more fine-grained scale (i.e., a hundred levels).

\myparagraph{The Impact of Assessors' Bias.}
Human bias is often reflected in manually labeled datasets and therefore in supervised systems that make use of such data. For example,
\citet{otterbacher2017competent} showed that human bias and stereotypes are reflected in search engine results.
In the context of crowdsourced relevance judgments, \citet{eickhoff2018cognitive} showed how common types of bias can impact the collected judgments and the results of Information Retrieval (IR) evaluation initiatives. Previous studies have found a positive correlation between cognitive skills (measured by means of the cognitive reflection test or CRT \cite{Frederick2005})
and the ability to identify true and false news \cite{pennycook2018falls,pennycook2019lazy,pennycook2019fighting}. 
In our work we collect assessors' background and bias data to then identify patterns in their assessment behaviors.

%% file: 3-method.tex
\section{Experimental Setting}\label{sec:exp}
In this section we first introduce the two datasets that we used in our experiments. We then present the task design we created and the different rating scales we used to collect truthfulness assessments from crowd workers.
In our experiments we use two \textit{datasets} (i.e., two sets of statements made by politicians), eight judgment \textit{collections} (i.e., three sets of crowd judgments per dataset and one expert judgment set per dataset), and three different judgment \textit{scales} (i.e., with three, six, and one hundred levels).

\subsection{Datasets} 
\begin{table}[tbp]
\centering
\caption{Example of statements in the \politifact and \abc datasets. 
}
\label{tab:statements}
\begin{adjustbox}{max width=0.47\textwidth}
\begin{tabular}{p{2.4cm}p{3.5cm}p{2cm}}
\toprule
 & \textbf{Statement} & \textbf{Speaker, Year}\\
 \midrule
\politifact \mbox{Label: \politifactfour} & ``Florida ranks first in the nation for access to free prekindergarten.'' & Rick Scott, 2014 \\ 
\addlinespace
\abc \mbox{Label: \abcone} & ``Scrapping the carbon tax means every household will be \$550 a year better off.'' & Tony Abbott, 2014 \\ 
\bottomrule
\end{tabular}
\end{adjustbox}

\end{table}

\myparagraph{\politifact.} This dataset (constructed by \citet{politifact}) contains \num{12800} statements given by US politicians with truthfulness labels produced by expert fact-checkers on a 6-level scale (detailed in Section~\ref{sect:scales}). In this work, we selected a subset of 20 statements for each truth level from the original dataset covering the time span 2007 to 2015. The sample includes statements by politicians belonging to the two main US parties (Democratic and Republican).

\myparagraph{\abc.} This dataset published by 
RMIT ABC Fact Check\footnote{\url{https://www.abc.net.au/news/factcheck/}} consists of 407 verified statements covering the time span 2013 to 2015.
To create this dataset, professional fact checkers seek expert opinions and collect evidence before a team makes a collective decision on how to label each claim.
To this aim, a fine-graded scale is used and verdicts are labeled as: `Correct', `Checks out', `Misleading', `Not the full story', `Overstated', `Wrong', among others. These verdicts are then grouped in a three-level scale: \texttt{Positive}, \texttt{In-Between}, and \texttt{Negative}. In our experiments, the latter three-level scale is used as ground truth. Our sample includes 60 randomly selected statements (20 statements for each truth level) by politicians belonging to the two main Australian parties (i.e., Liberal and Labor).
For both \politifact and \abc datasets, a balanced number of statements per class and per political party was included in the sample.
Table~\ref{tab:statements} shows an example of \politifact and \abc statements.

\subsection{Crowdsourcing Task Design}
We collect truthfulness judgments using the crowdsourcing platform Amazon Mechanical Turk (MTurk).\footnote{The experimental setup was reviewed and approved by the Human Research Ethics Committee at The University of Queensland.}
Each worker accepting our Human Intelligence Task (HIT) receives a unique \emph{input token}, which identifies uniquely both the MTurk HIT and the worker, and is redirected to an external website where to complete the task.
The task is designed as follows: in the first part the workers are asked to provide some details about their background, such as age, family income, political views, the party in which they identifies themselves, their opinion on building a wall along the southern border of United States, and on the need for environmental regulations to prevent climate change. 
Then, to assess their cognitive abilities, workers are asked to answer three modified Cognitive Reflection Test (CRT) questions, which are used to measure whether a person tends to overturn the incorrect ``intuitive'' response, and further reflect based on their own cognition to find the correct answer. Psychologist Shane Frederick firstly~\cite{Frederick2005} proposed the original version of the CRT test in 2005. These modified questions are: 
\begin{itemize}
\item If three farmers can plant three trees in three hours, how long would it take nine farmers to plant nine trees?[correct answer = 3 hours; intuitive answer = 9 hours]
\item Sean received both the 5th highest and the 5th lowest mark in the class. How many students are there in the class? [correct answer = 9 students; intuitive answer = 10 students]
\item In an athletics team, females are four times more likely to win a medal than males. This year the team has won 20 medals so far. How many of these have been won by males? [correct answer = 4 medals; intuitive answer = 5 medals]
\end{itemize}

After this initial survey, workers are asked to provide truthfulness values for 11 statements: 6 from \politifact, 3 from \abc, and 2 which serve as gold questions --one obviously true and the other obviously false-- written by the authors of this paper. All the \politifact statements we use come from the most frequent five \emph{contexts} (i.e., the circumstance or media in which the statement was said / written) available in the dataset; to avoid bias, we select a balanced amount of data from each context. 
To assess the truthfulness of statements, workers are presented with the following information about each  statement:
\begin{itemize}
    \item \emph{Statement}: the statement.
    \item \emph{Speaker}: the name and surname of whom said the statement.
    \item \emph{Year}: the year in which the statement was made.
\end{itemize}

We asked each worker to provide both the truthfulness level of the statement, and a \emph{URL} that serves both as justification for their judgment as well as a source of evidence for fact checking.
In order to avoid workers finding and using the original  expert labels (which are available on the Web) as primary source of evidence,
we ask workers to use a provided custom web search engine to look for supporting evidence.
The custom search engine uses the Bing Web Search API,\footnote{\url{https://azure.microsoft.com/services/cognitive-services/bing-web-search-api/}} and  filters out from the retrieved results any page from the websites that contain the  collection of expert judgments we used in our experiment.
After workers assess all 11 statements they can submit the HIT.
In order to increase the quality of collected data, we embedded the following quality check in the crowdsourcing task:
\begin{itemize}
    \item \emph{Gold Questions}: the worker must assign to the obviously false statement a truthfulness value lower than the one assigned to the obviously true statement.
    \item \emph{Time Spent}: the worker must spend at least two seconds on each statement and cognitive question.
\end{itemize}
We performed several small pilots of the task, and after measuring the time and effort taken to successfully complete it, we set the  HIT reward to \$1.5. This was computed based on the expected time to complete it and targeting to pay at least the US federal minimum wage of \$7.25  per hour.
Given the aims of the experiment, we publish the task allowing only US-based workers to participate. 
During the experiment, we logged all worker behaviors using Javascript code in the HIT that sends log messages to our server at each worker action (e.g., clicking on a truthfulness level, submitting a search query, selecting a URL, or moving to the next statement).
To avoid learning effects, we choose to allow each worker to complete only one of our HITs for only one experimental setting (i.e., one judgment scale).
Overall, not including pilot runs whose data was then discarded, we collected assessments for 120 (\politifact) $+$ 60 (\abc) = 180 statements each judged by 10 distinct workers.
We repeated this over 3 different assessment scales, so, in total, we collected \num{1800} (for each scale) * 3 = \num{5400} assessments. If we consider also the assessments of the two gold questions we embedded in the task, workers provided a total of \num{6600} assessments.

\subsection{Assessment Scales and Collections}\label{sect:scales}
In our experimental design we consider \emph{three} truthfulness scales and generated \emph{five} collections: two ground truths labeled by experts (for \politifact and \abc), and three created by means of our crowdsourcing experiments (\three, \six, and \onehundred):
\begin{itemize}
    \item \emph{\politifact}: uses a six-level  scale, with labels \politifactzero, \politifactone, \politifacttwo, \politifactthree, \politifactfour, and \politifactfive.
    \item \emph{\abc}: uses a three-level   scale, with labels \abczero, \abcone, and \abctwo.
    \item \emph{\three}: uses a three-level  scale, with the same labels as the \abc scale.  
    \item \emph{\six}: uses a six-level   scale, with the same labels as the \politifact scale, but replacing \politifactzero with \politifactlie (we felt that ``Lie'' would be more clear than the colloquial ``Pants on Fire'' expression).
    \item \emph{\onehundred}: uses a one-hundred-and-one level scale, with values in the $[0, 100]$ range.\footnote{The number of levels of this scale is 101 but we call it \onehundred for simplicity.}
\end{itemize}

The nature and usage of these scales deserve some discussion. 
The scales we use are made of different levels, i.e., categories, but they are not nominal scales: they would be nominal if such categories were independent, which is not the case because they are ordered. This can be seen immediately by considering, for example, that misclassifying a \politifactfive statement as \politifactfour is a smaller error than misclassifying it as \politifactthree. 
Indeed all of them are ordinal scales. However, they are not mere rankings, as the output of an information retrieval systems, since statements are assigned to categories, besides being ranked: given two statements having as ground truth, say, \politifactfive and \politifactfour respectively, misclassifying them as \politifactthree and \politifacttwo is an error, and it is a smaller one than misclassifying them as \politifactone and \politifactzero, but in all cases the original ranking has been preserved.
These scales are sometimes named Ordinal Categorical Scales \cite{agresti2010analysis}.

For ordinal categorical scales it cannot be assumed that the categories are equidistant. For example, a misclassification from \politifactzero to \politifactone cannot be assumed to be a smaller error than a misclassification from \politifacttwo to \politifactfive, in principle. Again, to be rigorous, taking the arithmetic mean to aggregate individual worker judgments for the same statement into a single label is not correct, since this would assume equidistant categories. On the contrary, the mode (called `majority vote' by the crowdsourcing community), which is the right aggregation function for nominal scales, even if correct, would discard important information. For example, the aggregation of four \politifactzero with six \politifactone judgments should be rather different from --and lower than-- six \politifactone and four \politifactfive, though the mode is the same. The orthodox correct aggregation function for this kind of scale is the median. 

However, the situation is not so clear cut. In the last example, the median would give the exact same result than the mode, thus discarding useful information. A reasonably defined ordinal categorical scale would feature labels which are approximately equidistant. This is particularly true for \onehundred, since: (i) we used numerical labels $[0, 100]$ for the categories, and (ii) the crowd workers had to use a slider to select a value. This makes \onehundred (at least) very similar to an interval scale, for which the usage of the mean is correct --and indeed it has been already used for \onehundred  \cite{Roitero:2018:FRS:3209978.3210052}. Also, in information retrieval we are well used to interpreting ordinal scales as interval ones (e.g., when we assign arbitrary gains in the NDCG effectiveness metrics) and/or (ab)using the arithmetic mean (e.g., when  we take the mean of ranks in the Mean Reciprocal Rank metric) \cite{Fuhr:2018}. In many practical cases, using the arithmetic mean turns out to be not only adequate but even more useful than the correct aggregation function~\cite{graham2015accurate, mathur2017sequence,aletras2017evaluating}. 
Even worse, there are no metrics for tasks defined on an ordinal categorical scales, like predicting the number of ``stars'' in a recommendation scenario. Sometimes accuracy is used, like in NTCIR-7 \cite{Kando-08}, but this is a metric for nominal scales; in some other cases, like in RepLab~2013~\cite{replab2013},  Reliability and Sensitivity~\cite{amigo2013general} are used, which consider only ranking information and no category membership; even metrics for interval scales, like Mean Average Error (MAE), have been used \cite{Ghosh-15}.

For these reasons, in the following 
we sometimes use the (aggregated) truthfulness labels expressed by workers as if they where on an interval scale. Another reason to do so is to treat the various scales in a homogeneous way, and thus use the same aggregation used for \onehundred also for \three, \six, \politifact, and \abc.
Accordingly, in the following we denote the labels of \abc and \three with \texttt{0}, \texttt{1}, and \texttt{2}, as if they where in the $[0, 2]$ range. Moreover, we denote the labels of \politifact and \six with \texttt{0},  \texttt{1}, \ldots, \texttt{5} --as if they where expressed in the $[0, 5]$ range. Finally,  we denote the truthfulness labels of \onehundred with \texttt{0},  \texttt{1}, $\ldots$, \texttt{100}.

We take these as working assumptions for our setting, and we leave an exhaustive study of the perceived distance between the truthfulness levels for future work; furthermore, we perform our initial analyses aggregating by means but then discuss the effects of using alternative aggregation functions in Section~\ref{sect:alternative-aggregation}.

%% file: 4-Analysis.tex
\section{Results}\label{sec:res}
\subsection{Worker Background and Behavior}\label{sect:worker-questionnaire}
About six hundred US resident\footnote{MTurk workers based in the US must provide evidence they are eligible to work.}
crowd workers participated to this study on MTurk.
Across all experiments, 
the majority of workers (46.33\%) are between 26 and 35 years old. Second, workers in our study are well-educated as more than 60.84\% of workers have a four years college degree at least. Third, around 67.66\% of  workers earn less than \$75,000 a year.

Nearly half (47.33\%) of the workers think their views are more democratic and only about 22.5\% of workers selected the Republican Party as voting preference. As for political views, Liberal and Moderate accounted for the most substantial proportion of workers, 29.5\% and 28.83\% respectively, while Very Conservative accounted for the least, only 5.67\%. In response to the border issue, 52.33\% of US-based workers  are against the construction of a wall on the southern border, while 36.5\% of the workers supported the building. For environmental protection, 80\% of the workers supported the government to strengthen environmental regulation to prevent climate change, while 11.33\% of the workers objected.

Based on the behavioral actions we logged as workers went through the HITs, Table \ref{tab:worker-behavior-data} shows the ratio of workers who completed the task, abandoned the task, and failed the quality checks.
Abandonment numbers are in line with previous studies \cite{8873609}. We can observe a higher failure and lower completion rate for \onehundred. This may show a slight lack of comfort for workers in using the most fine-grained scale.
We also logged worker behavior in terms of going back to previously seen statement (less than 5\% over all scales).

\begin{table}[tbp]
\caption{Worker behavior rates (percentage).}
 \label{tab:worker-behavior-data}
    \centering
    \begin{tabular}{l@{ }ccccccccc}
\toprule
& \textbf{Completion} & \textbf{Abandonment} & \textbf{Failure}\\
\midrule
\three & 35 & 53 & 12\\
\six & 33 & 52 & 14\\
\onehundred & 25 & 53 & 22\\
\bottomrule
\end{tabular}
\end{table}

\subsection{Crowdsourced  Score Distributions}\label{sect:score-distribution}

\begin{figure}[tbp]
  \centering
  \begin{tabular}{@{}c@{}c@{}c@{}}
    \includegraphics[width=.33\linewidth]{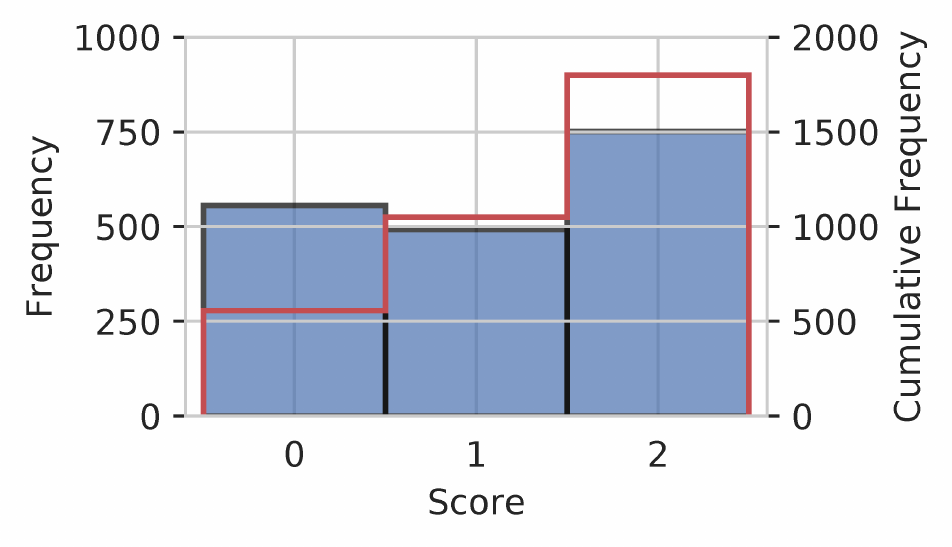}&
    \includegraphics[width=.33\linewidth]{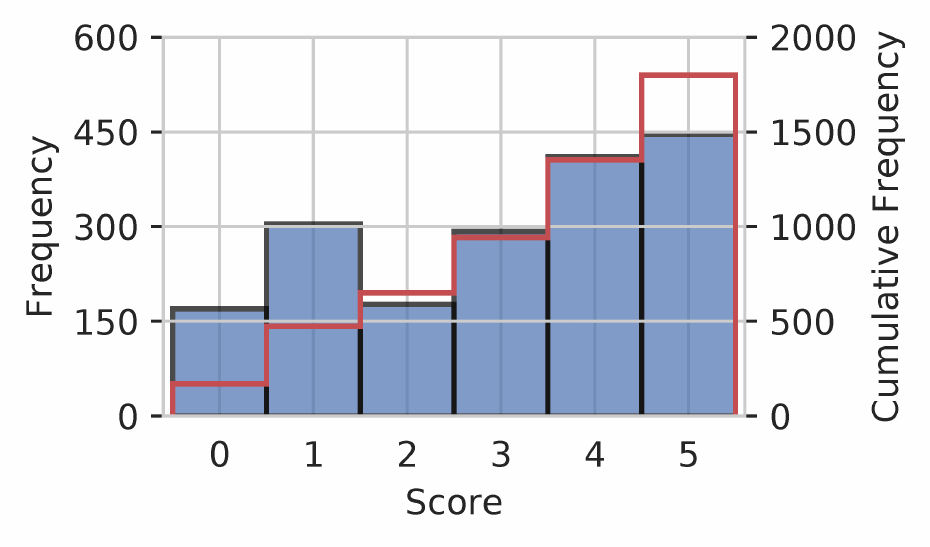}&
    \includegraphics[width=.33\linewidth]{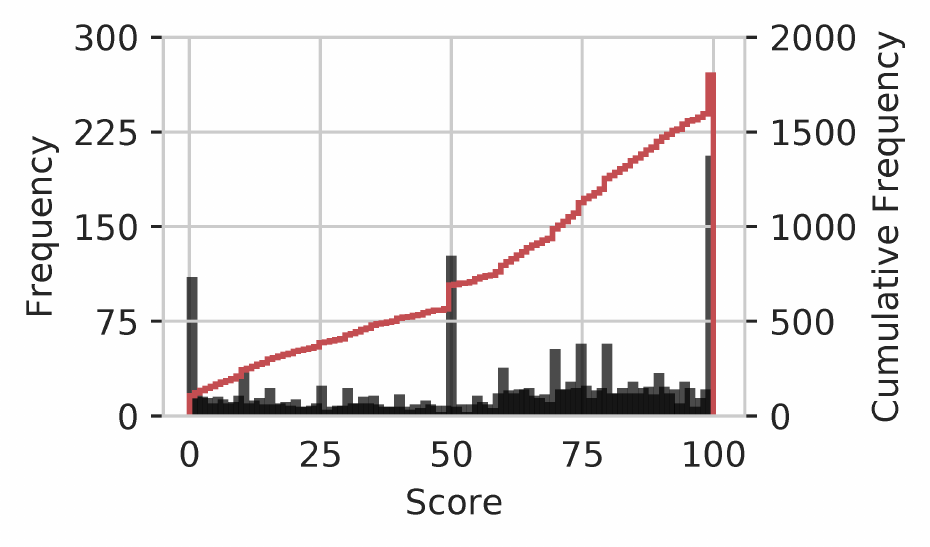}\\
    \includegraphics[width=.33\linewidth]{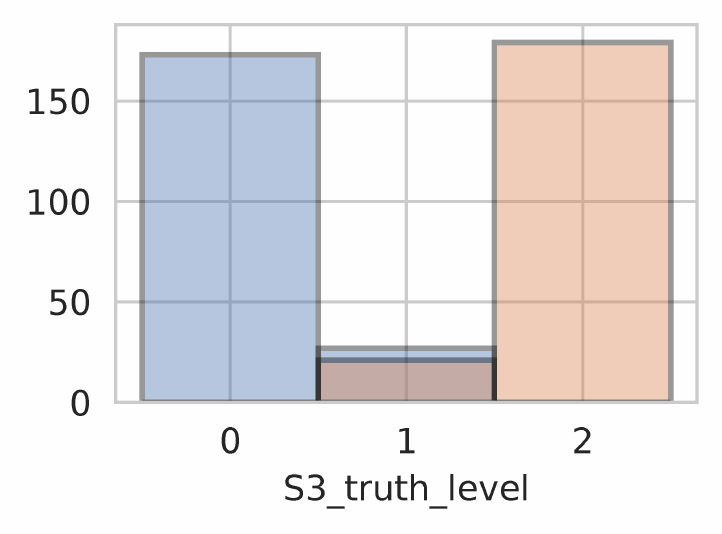}&
    \includegraphics[width=.33\linewidth]{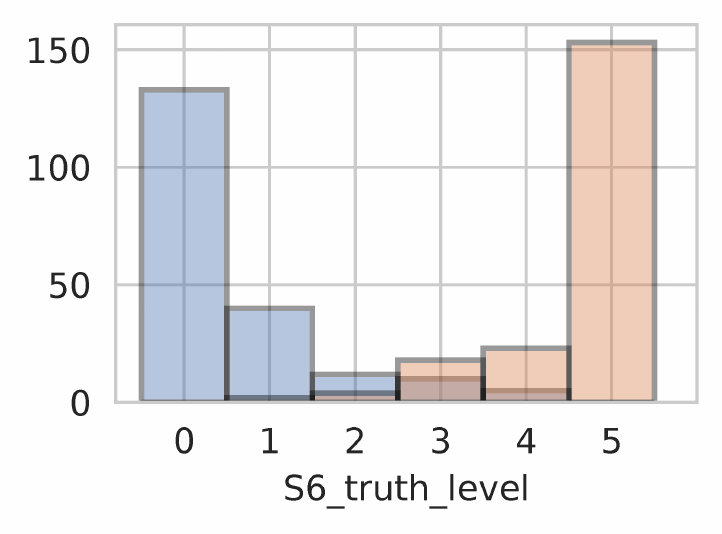}&
    \includegraphics[width=.33\linewidth]{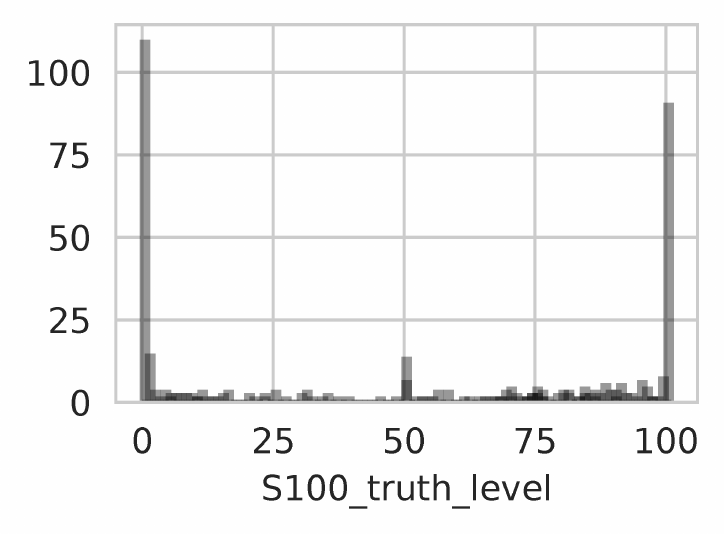}\\
    \includegraphics[width=.33\linewidth]{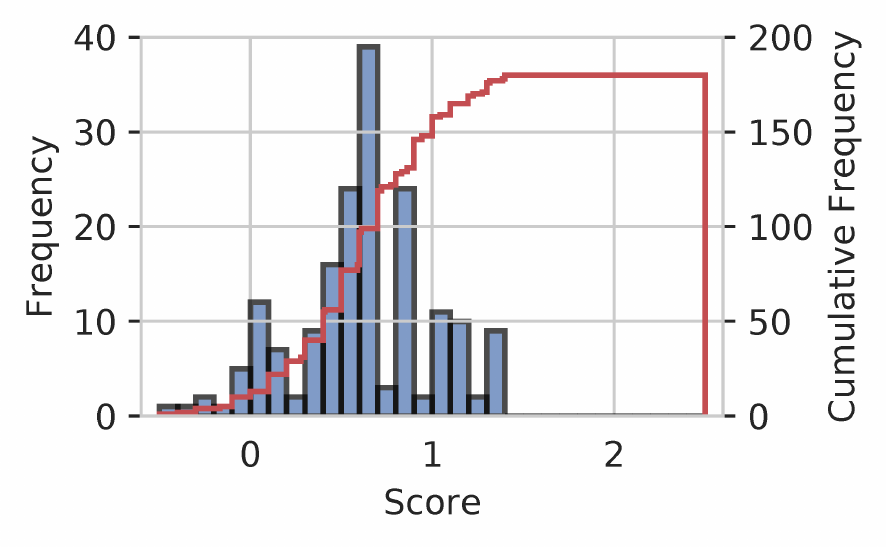}&
    \includegraphics[width=.33\linewidth]{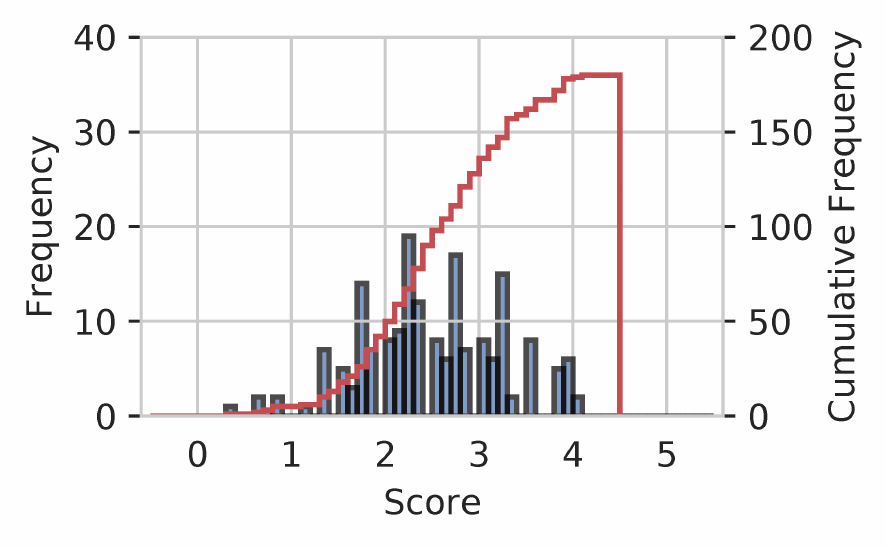}&
    \includegraphics[width=.33\linewidth]{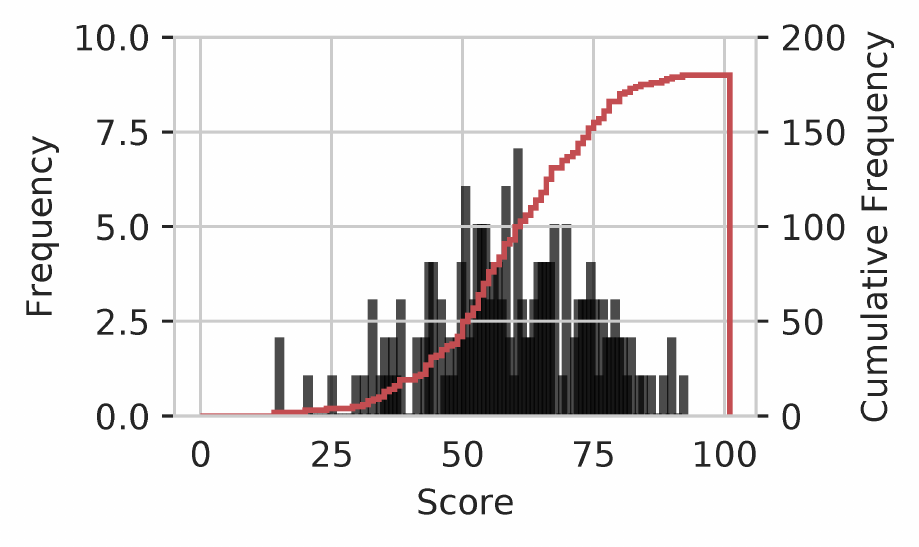}\\
    
  \end{tabular}
\caption{%
    From left to right: \three, \six, \onehundred.
    From top to bottom: individual scores distribution (first row),
    gold scores distribution (second row), and
    aggregated scores distribution (third row). 
}
  \label{fig:scores_distributions}
\end{figure}

This section presents an analysis of score distributions obtained by the crowd for the statements used in our experiment.
Figure~\ref{fig:scores_distributions} shows the distribution (and the cumulative distribution in red) for the individual scores provided by workers over the three crowd collections (i.e., \three, \six, and \onehundred) for \emph{all} the statements considered in the experiment; the behavior is similar across  \politifact and \abc statements (not shown because of space limitations).

The first row of the figure 
shows the raw score distributions for the three collections. As we can see from the plots, for all \three, \six, and \onehundred the distribution is skewed towards the right part of the scale representing higher truthfulness values; this can be also seen by looking at the cumulative distribution, which is steeper on the right hand side of the plots. 
It can also be seen that all three distributions are multimodal.
Looking at \onehundred we can  see a mild \emph{round number tendency}, that is, the tendency of workers to provide truthfulness scores which are multiple of 10 (35\% of \onehundred scores are multiple of 10; 23\% are 0, 50, or 100); such behavior was already noted by  \citet{Maddalena:2017:CRM:3026478.3002172}, \citet{RSDM:2018}, and \citet{Roitero:2018:FRS:3209978.3210052}. Also in this case, this behavior is consistent when considering separately \politifact and \abc documents (not shown).

We now turn to \emph{gold} scores distribution, i.e., the special statements \high and \low that we used to perform  quality checks during the task. 
The second row of Figure~\ref{fig:scores_distributions} shows the scores distribution for the three crowd collections we considered. As we can see from the plots, the large majority of workers 
(44\% for \low and 45\% for \high in \three, 
34\% for \low and 39\% for \high in \six, and 
27\% for \low and 24\% for \high in \onehundred) 
provided as truthfulness level for these gold statements the extreme value of the scales (respectively the lower bound of the scale for \low and the upper bound of the scale for \high). This can be interpreted as a signal that the gathered data is of good quality. However, some workers provided judgments inconsistent with the gold labels. 

Finally, we discuss \emph{aggregated} scores. The third row of Figure~\ref{fig:scores_distributions} shows the score  distributions of \three, \six, and \onehundred judgments aggregated by taking the \emph{average} of the 10 scores obtained independently for each statement. 
As we can see from the plots, the distribution of the aggregated scores for all \three, \six, and \onehundred  are similar, they are no longer multimodal, and they are roughly bell-shaped.  
It is worth noting that, while the aggregated scores for \three (bottom-left plot in Figure \ref{fig:scores_distributions}) are skewed towards lower/negative --i.e., \abczero and \abcone-- scores, \three and \onehundred (bottom-mid and bottom-right plots in Figure \ref{fig:scores_distributions}) are skewed to higher/positive scores.  
This shows how different scales are used differently by the crowd.
As expected \cite{Maddalena:2017:CRM:3026478.3002172,RSDM:2018,Roitero:2018:FRS:3209978.3210052}, for \onehundred the round number tendency effect also disappears when judgments from different workers are aggregated together.
In the following we discuss how  truthfulness scores gathered from the crowd compare to expert labels.

\subsection{External and Internal Agreement}\label{sect:agreement-experts}

We now discuss the external agreement between the crowd judgments and the expert labels as well as the internal agreement among workers, addressing RQ2.

\begin{figure*}[tbp]
  \centering
  \begin{tabular}{@{}c@{}c@{}c@{}}
    \includegraphics[width=.33\linewidth]{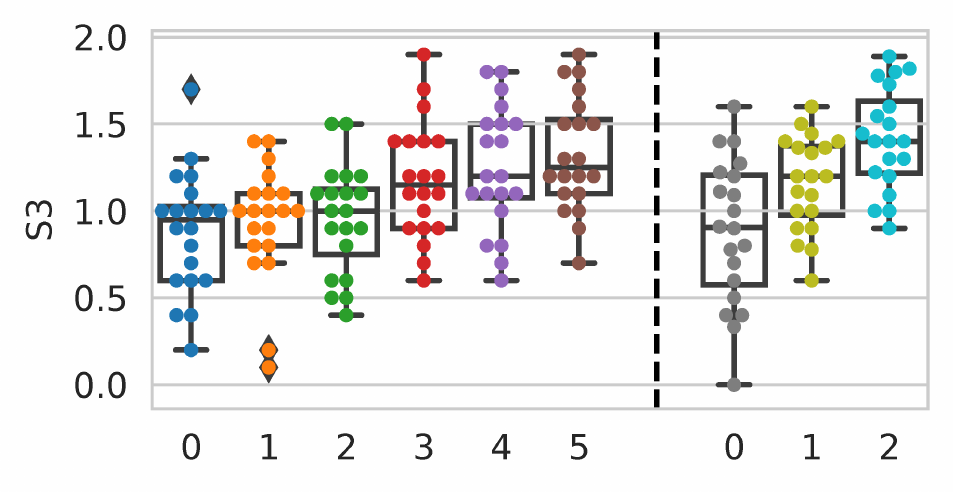}&
    \includegraphics[width=.33\linewidth]{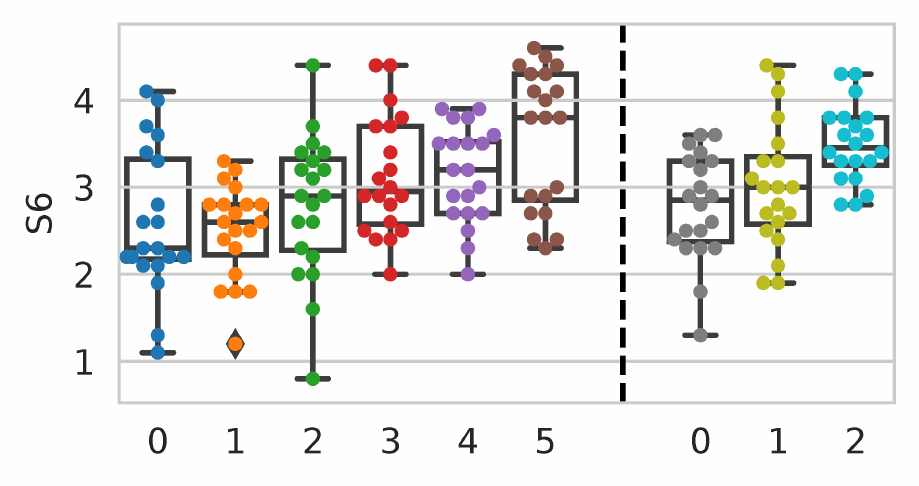}&
    \includegraphics[width=.33\linewidth]{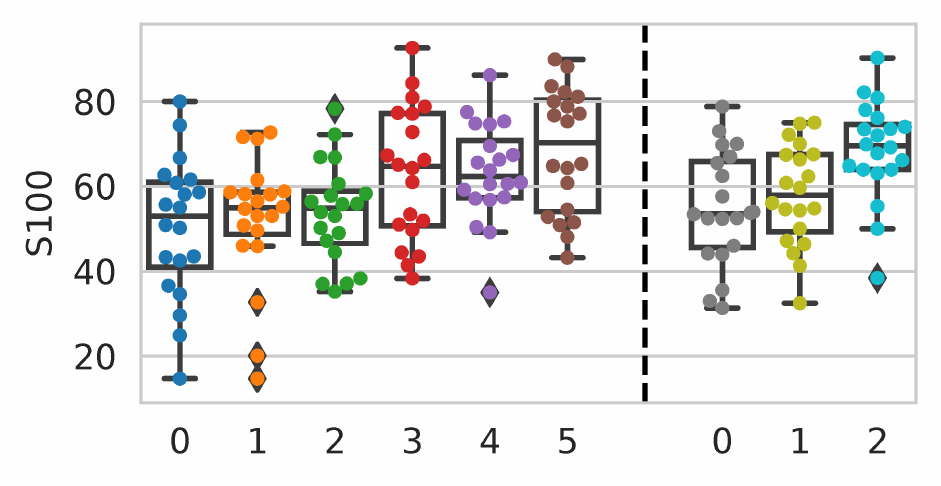}\\
  \end{tabular}
\caption{%
    From left to right: \three, \six, \onehundred;
    agreement with \politifact and \abc, separated by the vertical dashed line. 
  \label{fig:scale-comparison-ground}}
\end{figure*}

\myparagraph{External Agreement.}
Figure~\ref{fig:scale-comparison-ground} shows the agreement between the aggregated crowd judgments and the ground truth (i.e., the expert labels provided for \politifact and \abc), for the \three, \six, and \onehundred crowd collections. As we can see from the plots, the behavior over all the three scales is similar, both on \politifact and \abc statements.
If we focus on \politifact documents (shown in the first panel of each plot), we can see that it is always the case that the \texttt{0} and \texttt{1} boxplot are very similar. This can point out a difficulty from workers to distinguish between the \politifactzero and \politifactone labels. The same behavior, even if less evident, is present between the \politifactone and \politifacttwo labels; this behavior is consistent across all the scales.
On the contrary, if we focus on the rest of the  \politifact labels, and on the \abc ones, we can see that the median lines of each boxplot are increasing while going towards labels representing higher truth values (i.e., going towards the right hand side of each chart), indicating that workers have higher agreement with the ground truth for those labels. Again, this behavior is consistent and similar for all the \three, \six, and \onehundred scales.

We measured the statistical significance of the differences between the ratings aggregated by mean for categories of the \six, \three, and \onehundred collections according to the Mann-Whitney rank test and the t-test. 
Concerning \abc, adjacent categories are significantly different in 5 cases out of 12, while the difference between non adjacent categories are all significant to the p$<.01$ level.
Concerning \politifact, the differences between the ratings aggregated by mean for adjacent categories and not adjacent ones by distance of 2 (e.g., \texttt{0} and \texttt{2})  are never significant with only one exception (distance 2);
differences for not adjacent categories of distance 3 are significant in 4/18 cases, and
differences for categories of distance 4 are significant in 5/12 of the cases.
Finally, categories of distance 5 (i.e., \texttt{0} and \texttt{5}) are significant in 4/6 cases.
Although there is some signal, it is clear that the answer to RQ1 cannot be positive on the basis of this results. We will come back on this in Section \ref{sect:merge}.

\begin{figure}[tbp]
  \centering
  \begin{tabular}{@{}c@{}c@{}}
    \includegraphics[width=.48\linewidth]{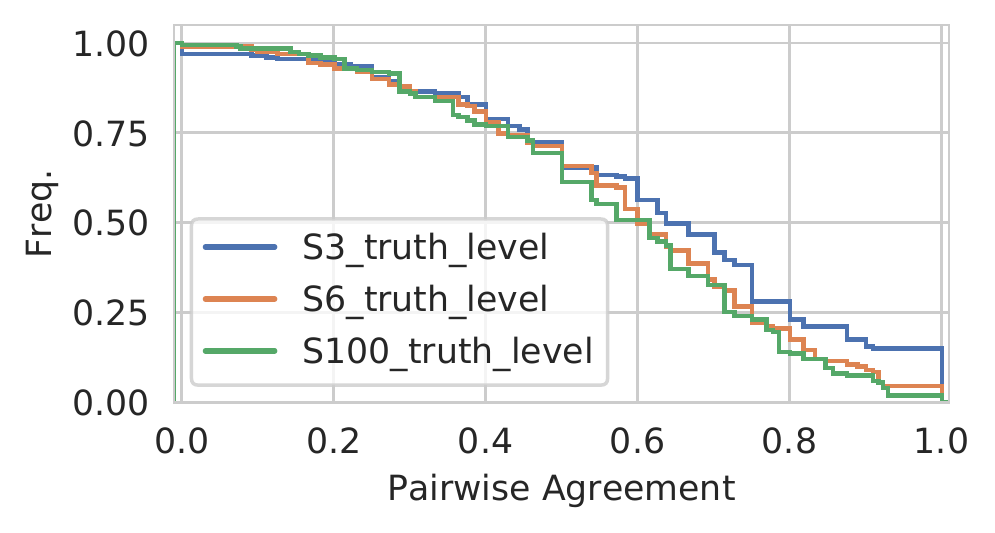} & 
    \includegraphics[width=.48\linewidth]{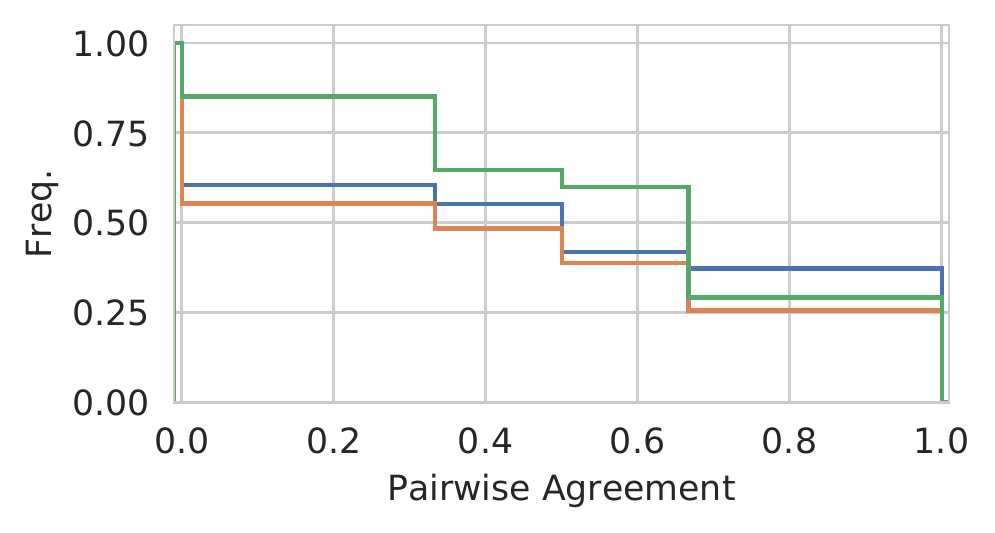} \\
  \end{tabular}
\caption{%
    HIT pairwise agreement, relative frequency (\politifact on the left, \abc on the right).
  \label{fig:pairwise-unit-agreement}}
\end{figure}

\begin{figure}[tbp]
  \centering
  \begin{tabular}{@{}c@{}c@{}c@{}}
    \includegraphics[width=.33\linewidth]{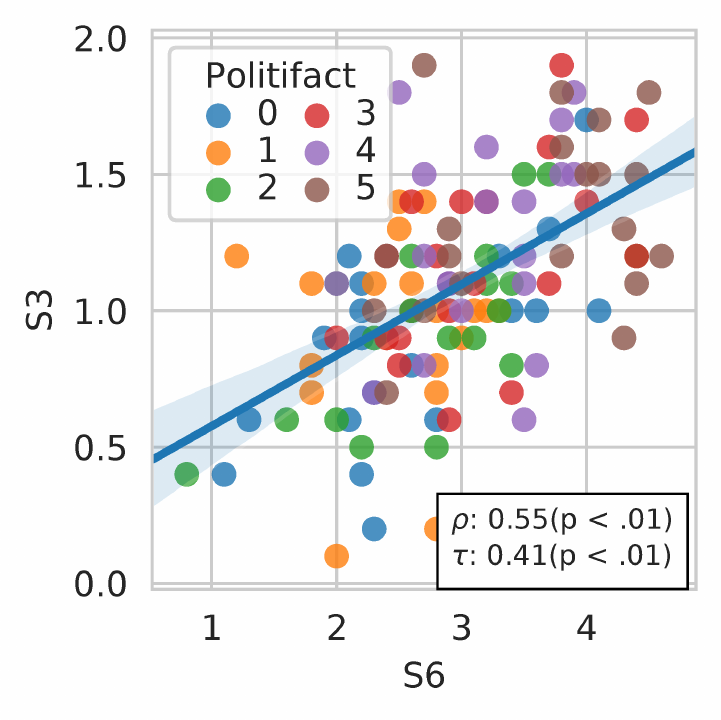}&
    \includegraphics[width=.33\linewidth]{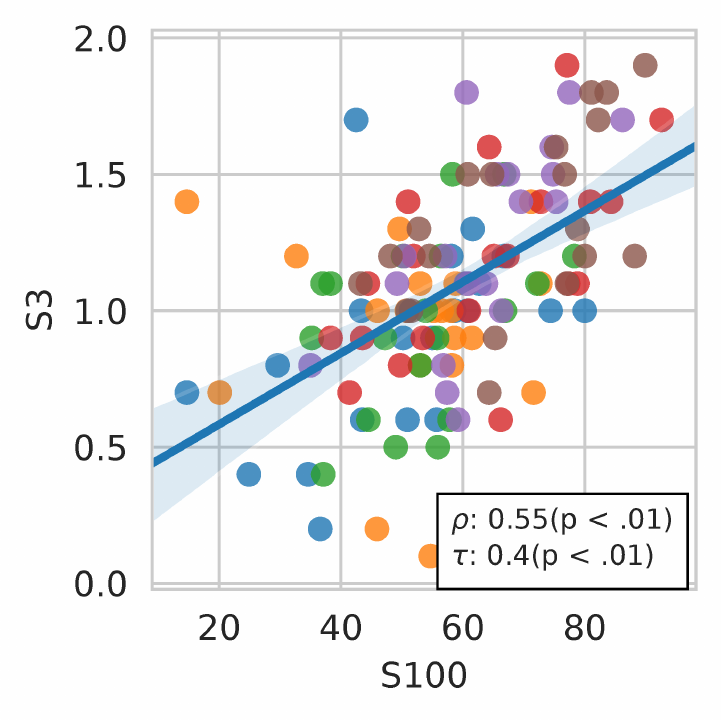}&
    \includegraphics[width=.33\linewidth]{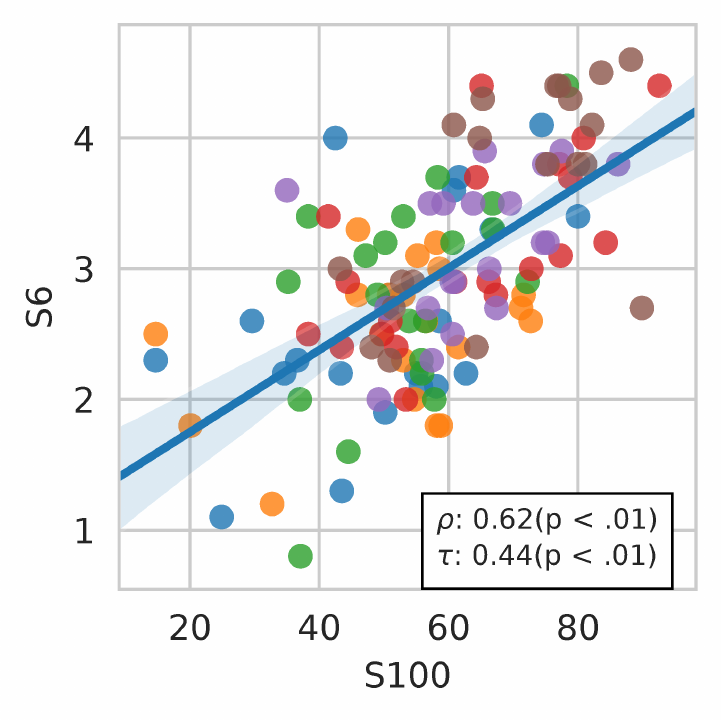}\\
    \includegraphics[width=.33\linewidth]{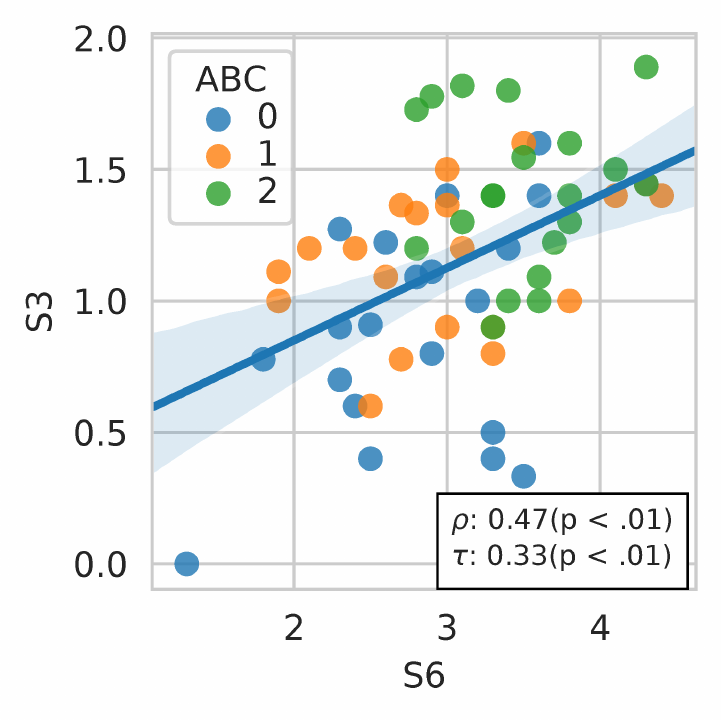}&
    \includegraphics[width=.33\linewidth]{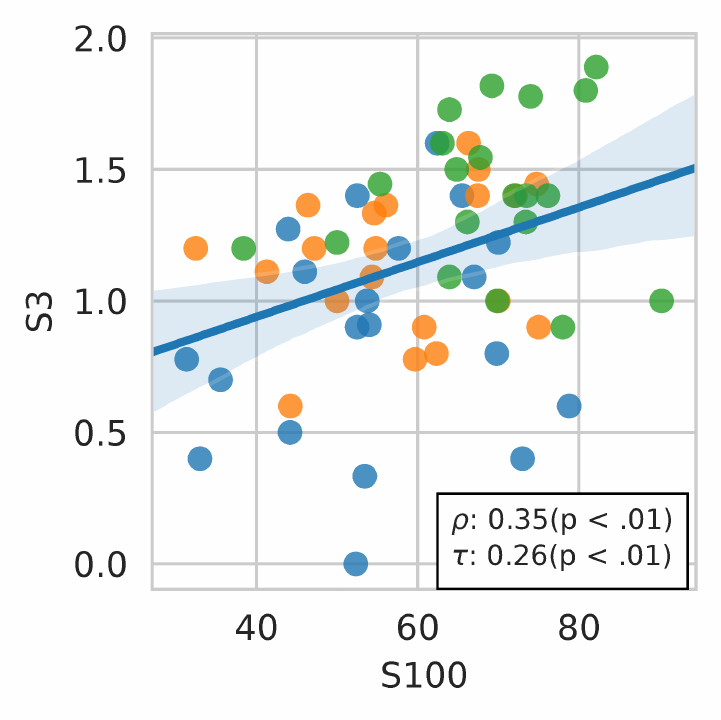}&
    \includegraphics[width=.33\linewidth]{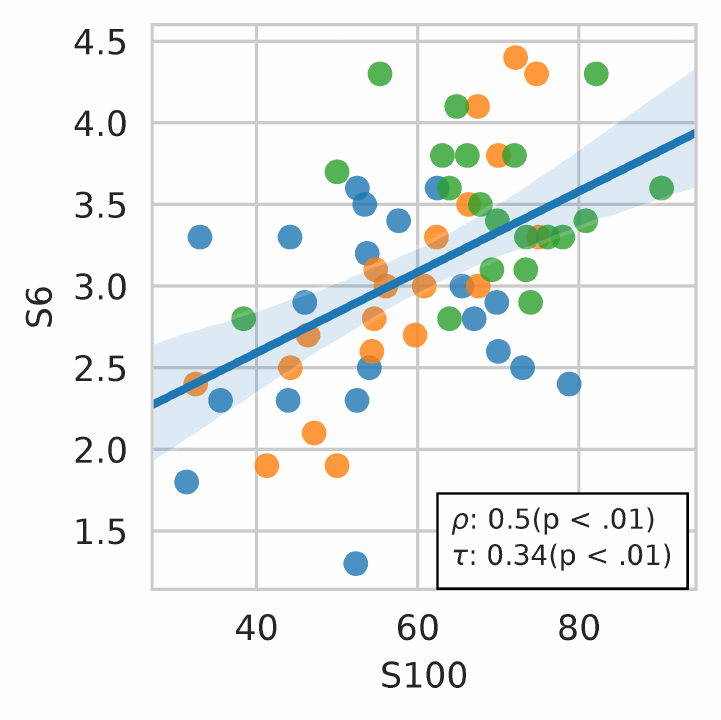}\\
  \end{tabular}
\caption{%
    Agreement between scales with a breakdown on \politifact statements (first row), and
    agreement between scales with a breakdown on \abc statements (second row).  From left to right: \six vs. \three, \onehundred vs. \three, and \onehundred vs. \six.  
  \label{fig:scale-comparisons}}
\end{figure}

We now turn on inspecting the agreement between the workers and the ground truth by looking at each HIT.
To do so, we computed, for all the \three, \six, and \onehundred collections the pairwise agreement \cite{10.1145/3121050.3121060} between the truthfulness scores expressed by workers and the ground truth labels, with a breakdown over \politifact and \abc statements.
We considered a slightly modified version of the pairwise agreement measure defined by \citet{10.1145/3121050.3121060}: in the attempt to make the pairwise agreement measure fully comparable across the different scales, we removed all the ties. 
Intuitively, pairwise agreement as described in \citet{10.1145/3121050.3121060} measures the fraction of pairs in agreement between a ``ground truth'' scale and a ``crowd'' scale.
Specifically, a pair of crowd judgments (crowd-judgment$_1$, crowd-judgment$_2$) is considered to be in agreement if crowd-judgment$_1$ $\leq$ crowd-judgment$_2$ and the ground truth for crowd-judgment$_1$ is $<$ the ground truth for crowd-judgment$_2$. In our measurement\footnote{The code used to compute the pairwise agreement as defined by us can be found at \url{https://github.com/KevinRoitero/PairwiseAgreement}.} we removed all the ties (i.e., crowd-judgment$_1$ $=$ crowd-judgment$_2$), and we used $<$ in place of $\leq$. 
Figure~\ref{fig:pairwise-unit-agreement} shows the CCDF (Complementary Cumulative Distribution Function) of the relative frequencies of the HIT agreement. As we can see from the charts, the \three, \six, and \onehundred scales show a very similar level of external agreement; such behavior is consistent across the \politifact and \abc datasets. Again, this result confirms that all the considered scales present a similar level of external agreement with the ground truth, with the only exception of \onehundred  for the \abc dataset: this is probably due to the treatment of ties in the measure, that removes a different number of units for the three scales.

\myparagraph{Internal Agreement.}
We now turn to  investigate the internal agreement (i.e., the agreement measured among the workers themselves), and in particular we also compare workers using different scales. 
We computed a metric used to measure the level of agreement in a dataset, the Krippendorff's $\alpha$ \cite{krippendorff2011computing} coefficient. 
All $\alpha$ values within each of the three scales \three, \six, \onehundred and on both \politifact and \abc collections are in the $0.066$--$0.131$ range.
These results show that there is a rather low agreement among the workers \cite{krippendorff2011computing, checco2017let}.

To further investigate if the low agreement  we found depends on the specific scale used to label the statements, we  also performed all the possible transformations of judgments from one scale to another, following the methodology described by \citet{scale}. 
Figure~\ref{fig:scale-comparisons} shows the scatterplots, as well as the correlations, between the different scales on the \politifact and \abc statements. As we can see from the plots, the correlation values are around $\rho=0.55$--$0.6$ for \politifact and $\rho=0.35$--$0.5$ for \abc, for all the scales. The rank correlation coefficient $\tau$ is around $\tau=0.4$ for \politifact and $\tau=0.3$ for \abc. These values indicate a low correlation between all the scales; this is an indication that the same statements on different scales tend to be judged differently, both when considering their absolute value (i.e., $\rho$) and their relative ordering (i.e., $\tau$). 

\begin{figure}[tbp]
  \centering
  \begin{tabular}{@{}c@{}c@{}c@{}}
    \includegraphics[width=.33\linewidth]{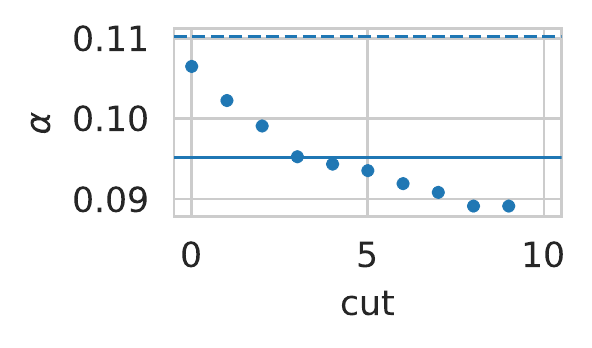} & 
    \includegraphics[width=.33\linewidth]{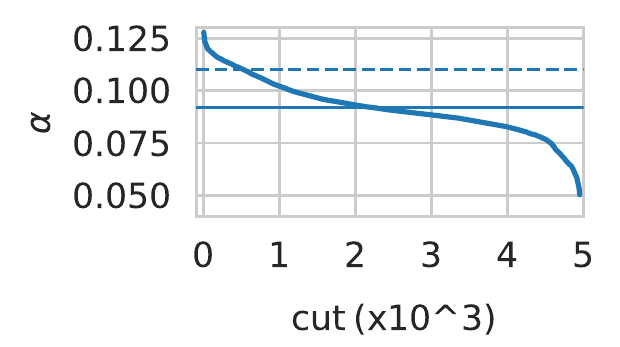} & 
    \includegraphics[width=.33\linewidth]{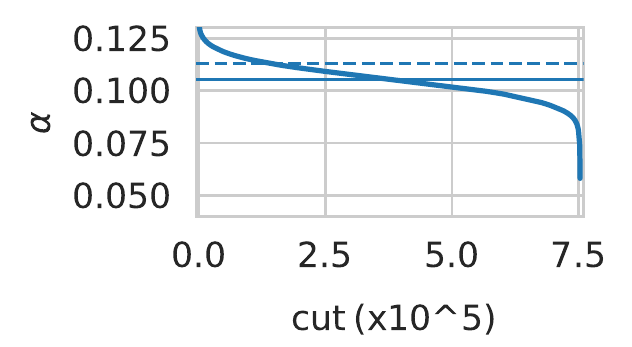}
  \end{tabular}
\caption{%
    From left to right: 
    \six cut into \three,
    \onehundred cut into \three, and
    \onehundred cut into \six (1\% stratified sampling), 
    cuts sorted by decreasing $\alpha$ values.
  \label{fig:alpha-cuts}}
\end{figure}

Figure~\ref{fig:alpha-cuts} shows the distribution of the $\alpha$ values when transforming one scale into another.\footnote{Note that the total number of possible cuts from \onehundred to \six is 75,287,520, thus we selected a sub-sample of all the possible cuts. We tried both \emph{stratified} and \emph{random} sampling, getting indistinguishable results.}
The dotted horizontal line in the plot represents $\alpha$ on the original dataset, the dashed line is the mean value of the (sampled) distribution.
As we can see from the plots, values on the y-axis are very concentrated and all $\alpha$ values are close to zero ($[0,0.15]$ range). 
Thus, we can conclude that across all collections there is low level of internal agreement among workers, both within the same scale and across different scales.

\subsection{Alternative Aggregation Functions}\label{sect:alternative-aggregation}

\begin{figure}[tbp]
  \centering
  \begin{tabular}{@{}c@{}c@{}c@{}}
    \includegraphics[width=.33\linewidth]{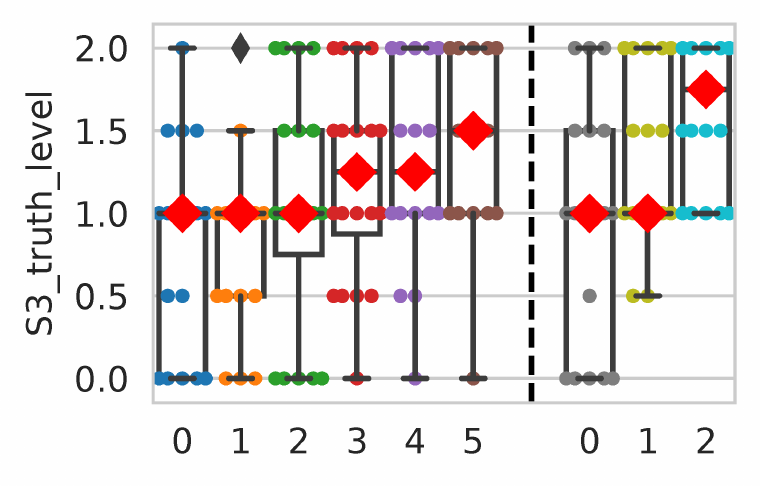}&
    \includegraphics[width=.33\linewidth]{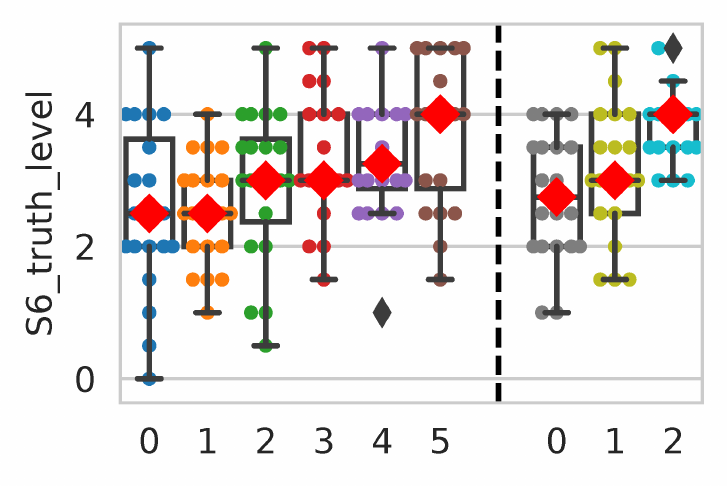}&
    \includegraphics[width=.33\linewidth]{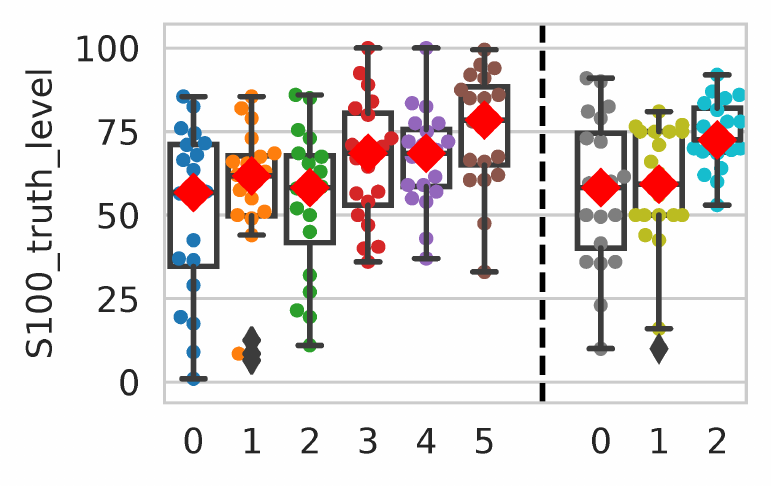}
  \end{tabular}
\caption{%
From left to right: \three, \six, \onehundred; agreement with ground truth.  Aggregation function: median (highlighted by the red diamond). Compare with Figure~\ref{fig:scale-comparison-ground}.
  \label{fig:alternative-aggregation-median}}
\end{figure}

\begin{figure}[tbp]
  \centering
  \begin{tabular}{@{}c@{}c@{}}
    \includegraphics[width=.30\linewidth]{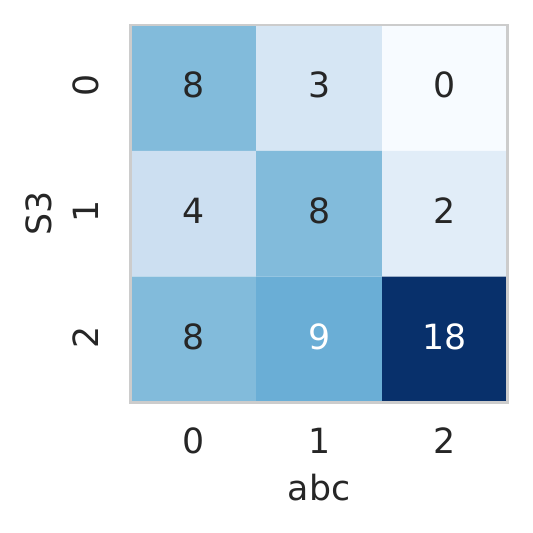}&
    \includegraphics[width=.49\linewidth]{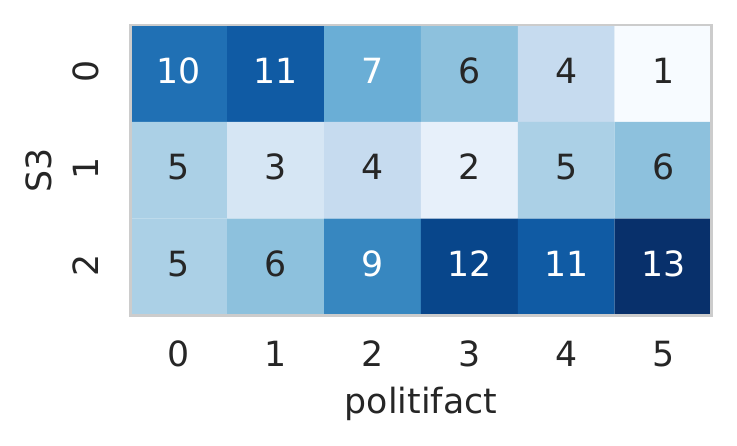}\\
    \includegraphics[width=.30\linewidth]{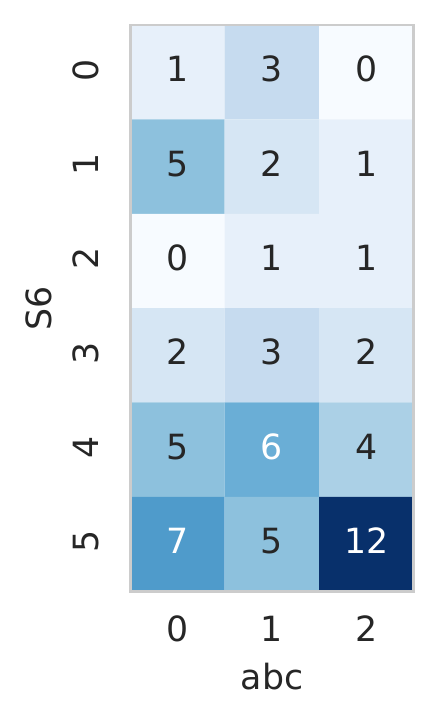}&
    \includegraphics[width=.49\linewidth]{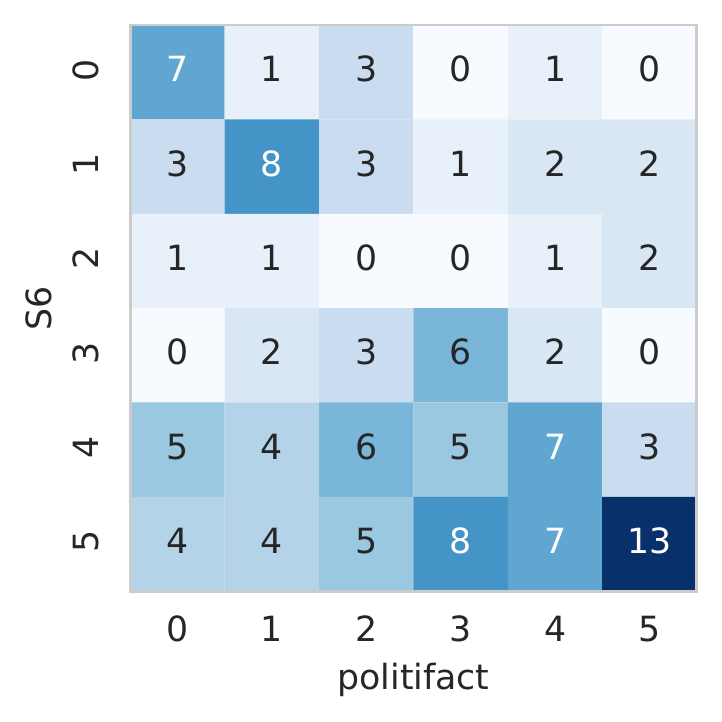}
  \end{tabular}
\caption{%
    Agreement between \three (first row) and \six (second row), and 
    \abc (left) and \politifact (right). Aggregation function: mode.
  \label{fig:alternative-aggregration}}
\end{figure}

As anticipated in Section~\ref{sect:scales}, we also study the effect of aggregation functions alternative to the arithmetic mean. 
Figure~\ref{fig:alternative-aggregation-median} shows the results of using the \emph{median}: in this case the final truthfulness value for each statement has been computed by considering the median of the scores expressed by the workers. By comparing the charts to those in Figure~\ref{fig:scale-comparison-ground} it is clear that the median produces worst results.
The heatmaps in Figure~\ref{fig:alternative-aggregration} show the results  of using the \emph{majority vote} (i.e., the mode) as the alternative aggregation function. The mode is  more difficult to compare with the mean, but it is again clear that the overall quality is rather low: although the squares around the diagonal tend to be darker and contain higher values, there are many exceptions. These are mainly in the lower-left corners, indicating false positives, i.e., statements whose truth value is over-evaluated by the crowd; this tendency to false positives is absent with the mean (see Figure~\ref{fig:scale-comparison-ground}). Overall these results confirm that the choice of mean as aggregation function seems the most effective.

\subsection{Merging Assessment Levels}\label{sect:merge}

Given the result presented so far in Sections~\ref{sect:agreement-experts} and~\ref{sect:alternative-aggregation} (especially Figures~\ref{fig:scale-comparison-ground}, \ref{fig:alternative-aggregation-median}, and~\ref{fig:alternative-aggregration}, but also the rather low agreement and correlation values), the answer to RQ1 cannot be completely positive. There is a clear signal that aggregated values resemble the ground truth, but there are also several exceptions and statements that are mis-judged. However, there are some further considerations that can be made. First, results seem better for \abc  than \politifact. 
Second, it is not clear which specific scale should be used. The two expert collections used as ground truth use different scales, and we have experimented with the crowd on \three, \six, and \onehundred. Also, comparisons across different scales are possible, as we have shown above. Finally, a binary choice (true/false) seems also meaningful, and in many real applications it is what may really be needed.
Third, the above mentioned possible
confusion between \politifactzero and \politifactone suggest that these two categories could be fruitfully merged. This has 
been  done, for example, by \citet{tchechmedjiev2019claimskg}. 

\begin{figure}[tbp]
  \centering
  \begin{tabular}{@{}c@{}c@{}c@{}c@{}c@{}c@{}}
    \includegraphics[width=.16\linewidth]{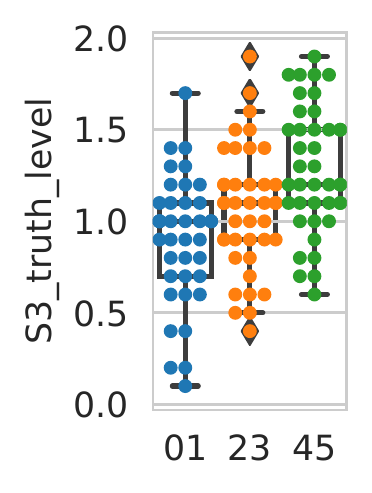}&
    \includegraphics[width=.16\linewidth]{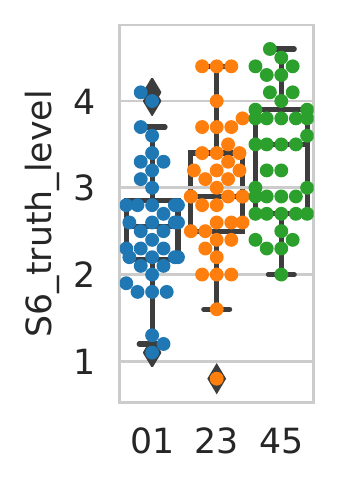}&
    \includegraphics[width=.16\linewidth]{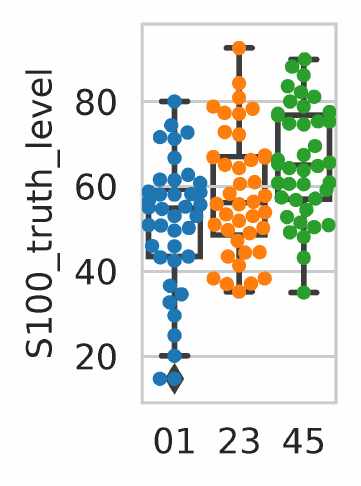}&
      \includegraphics[width=.16\linewidth]{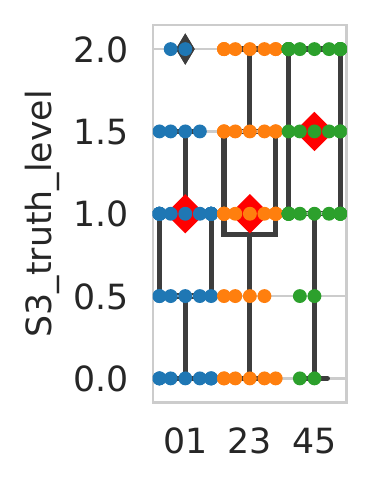}&
    \includegraphics[width=.16\linewidth]{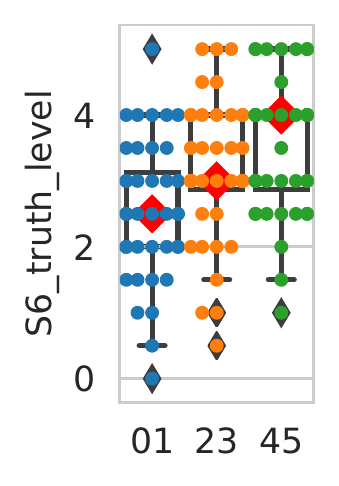}&
    \includegraphics[width=.16\linewidth]{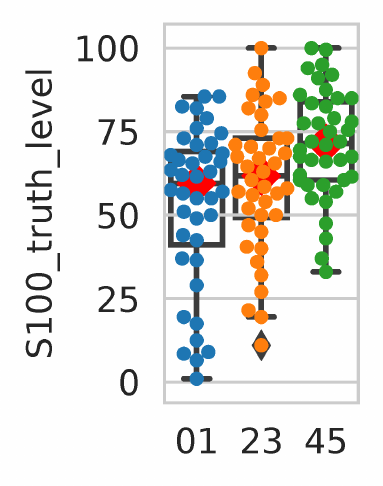}\\
    
      \includegraphics[width=.16\linewidth]{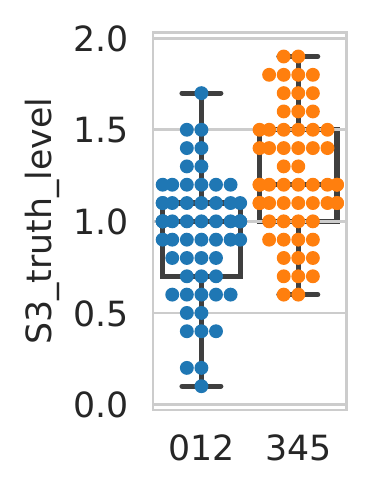}&
    \includegraphics[width=.16\linewidth]{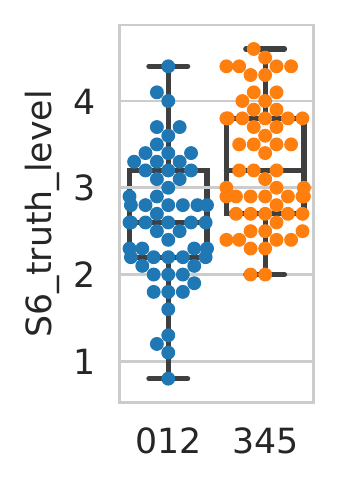}&
    \includegraphics[width=.16\linewidth]{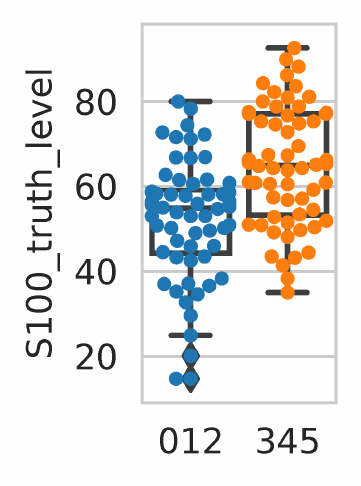}&
      \includegraphics[width=.16\linewidth]{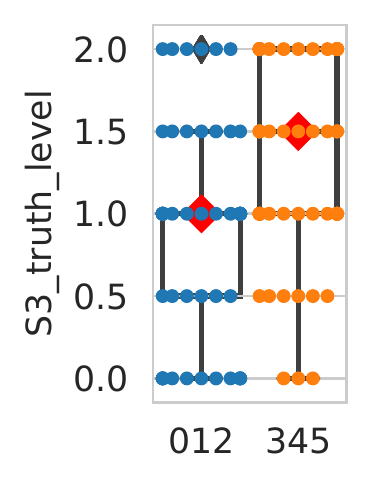}&
    \includegraphics[width=.16\linewidth]{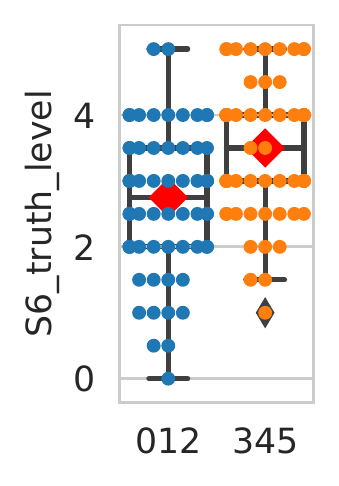}&
    \includegraphics[width=.16\linewidth]{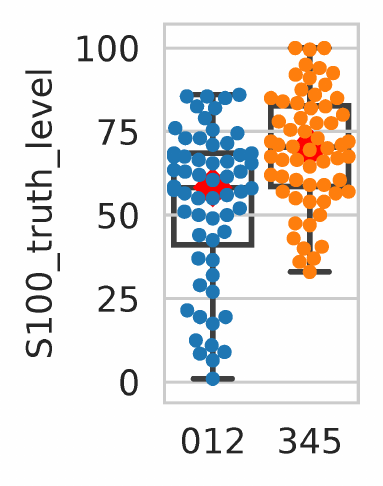}
  \end{tabular}
\caption{%
Agreement with ground truth for merged categories for \politifact. From the left:  mean for the three scales  \three, \six, \onehundred and then median for the same scales. From top to bottom: three  and two resulting categories. The median  is highlighted by the red diamond.
  \label{fig:alternative-aggregation-binned-3}}
\end{figure}

All these remarks suggest to attempt some grouping of adjacent categories, to check if  by looking at the data on a more coarse-grained ground truth the results improve. Therefore, we group the six \politifact categories into either three (i.e., \texttt{01}, \texttt{23}, and \texttt{45}) or two (i.e., \texttt{012} and \texttt{345}) resulting ones, adopting the approach discussed by \citet{scale}. 
Figure~\ref{fig:alternative-aggregation-binned-3} shows the results. The agreement with the ground truth can now be seen more clearly. The boxplots also seem quite well separated, especially when using the mean (the first three charts on the left). This is confirmed by analyses of statistical significance: all the differences in the boxplots on the bottom row are statistically significant at the p$<.01$ level for both t-test and Mann–Whitney, both with Bonferroni correction; the same holds for all the differences between \texttt{01} and \texttt{45} (the not adjacent categories) in the first row; for the other cases, i.e., concerning the adjacent categories,  further statistical significance is found at the  p<$.05$  level in 8 out of 24 possible cases. 

These results are much stronger than the previous ones: we can now state that the crowd is able to single out true from false statements with good accuracy; for statements with an intermediate degree of truthfulness/falsehood the accuracy is lower.

\subsection{Sources of Evidence}\label{sect:justification-distribution}

\begin{table*}[tbp]
\caption{Websites from which workers chose URLs to justify their judgments without considering gold questions for \three, \six, and \onehundred (percentage). Only websites with percentage $\geq 1\%$ are shown.
    \label{tab:justification-distribution-sheet}}
    \centering
    \begin{tabular}{@{}l@{ }c@{ }c@{ }c@{ }c@{ }c@{ }c@{ }c@{ }c@{ }c@{ }c@{ }c@{ }c@{ }c@{ }c@{ }c@{ }c@{ }c@{ }c@{ }c@{ }@{}}
\toprule
&  \footnotesize{Wikipedia} &   \footnotesize{Youtube} &   \footnotesize{The} &   \footnotesize{Factcheck} &  \footnotesize{Smh} &   \footnotesize{Cleveland} &   \footnotesize{Washington} &   \footnotesize{News} &   \footnotesize{Blogspot} &   \footnotesize{On the} &   \footnotesize{Quizlet} &   \footnotesize{NY} &   \footnotesize{CBS} &   \footnotesize{Forbes} &   \footnotesize{House} &   \footnotesize{Madison} &   \footnotesize{The} &   \footnotesize{Milwaukee} &   \footnotesize{Yahoo} \tabularnewline[-.5em]

 &  &   &   \footnotesize{Guardian} &    &   &    &  \footnotesize{Post}  &  &    &   \footnotesize{Issues} &    &  \footnotesize{Times} &   \footnotesize{News} &    &  & &   \footnotesize{Australian} &  \footnotesize{Journal}  &   \\
\midrule
\three  &      17 &    13 &        11 &       8 &  6 &       6 &            6 &  5 &      5 &         4 &     4 &     3 &      3 &    3 &   3 & 3 &           0 &      0 &   0 \\
\six   &      19 &    13 &        12 &       8 &  3 &       6 &            6 &  4 &      6 &         0 &     3 &     3 &      4 &   3 &   3 &     0 &           4 &      3 &   0 \\
\onehundred &      23 &    12 &        13 &       9 &  6 &       5 &            6 &  5 &      5 &         0 &     4 &     0 &      5 &    0 &   3 &     0 &           0 &      0 &   3 \\
\bottomrule
\end{tabular}
\end{table*}

Table~\ref{tab:justification-distribution-sheet} shows the distribution of websites used by workers to justify the truthfulness label they chose for each statement.
As we can see, the most used sources are, for all the scales, ``Wikipedia'' and ``YouTube'', followed by popular news websites such as ``The Guardian'' and  ``The Washington Post''. Furthermore, we can see that among the most popular sources there is one fact checking website (i.e., FactCheck). Noting that we intentionally removed \url{abc.com.au} and \url{politifact.com} URLs from those which could be selected, this shows that workers, supported by the search engine, tend to identify trustworthy information sources to support their judgment decisions.

\begin{table}[tbp]
\caption{Distribution of the ranks in search results for the URLs chosen by workers in \three, \six,  and \onehundred (percentage).}
 \label{tab:justification-distribution-rank}
    \centering
    \begin{tabular}{l@{ }ccccccccccc}
\toprule
& \textbf{1} & \textbf{2} & \textbf{3} & \textbf{4} & \textbf{5} & \textbf{6} & \textbf{7} & \textbf{8} & \textbf{9} & \textbf{10} & \textbf{11}\\
\midrule
\three          & 17 & 12 & 13 & 14 & 12 & 9  & 7 & 6 & 4 & 3 & 1\\
\six            & 13 & 13 & 16 & 12 & 11 & 9  & 8 & 7 & 5 & 4 & 1\\
\onehundred     & 15 & 15 & 15 & 12 & 8  & 12 & 7 & 6 & 5 & 2 & 1\\
\midrule
avg             & 15 & 13 & 15 & 11 & 10 & 10 & 7 & 6 & 5 & 3 & 1 \\
\bottomrule
\end{tabular}
\end{table}

Table~\ref{tab:justification-distribution-rank} 
shows the distribution of the ranks within the search engine results of the URLs chosen by workers to justify their judgments (without considering the gold questions), for \three, \six and \onehundred.
As we can see from the table, the majority of workers tend to click on the first results shown by the search engine, as expected \cite{kelly2015how,craswell2008experimental,joachims2017accurately}.
Nevertheless, the results also show that workers explore the first ten documents as ranked by the search engine and do not simply click on the first returned URL, thus putting some effort to find a reliable source and/or justification.
Finally, we note that over all the scales, all the workers stopped at the first page of results as returned by the search engine, and no one investigated search results with rank greater than 10.

\subsection{Effect of Worker  Background and Bias} \label{sect:worker-bias}

\begin{table}[tbp]
\centering
\caption{Correlation between Cognitive Reflection Test (CRT) performance and z-scores for each scale level and the correlation between worker age and z-scores.
}
\label{tab:Correlation }
\begin{threeparttable}

\begin{tabular}{llSS}
\toprule
{\textbf{Dataset}}          & {\begin{tabular}[c]{@{}c@{}}\textbf{Correlation} \\ \textbf{with}\end{tabular}}            & {\textbf{Age}}                        & {\begin{tabular}[c]{@{}c@{}}\textbf{CRT} \\ \textbf{Performance}\end{tabular}} \\ \midrule
\multirow{4}{*}{\politifact}                                          & Pants-on-fire & -0.038                     & -0.098$^{\ast}$                                                       \\
                                                                     & False         & -0.022                      & -.072                                      \\
                                                                     & True          & .127$^{\ast} $ & 0.062                                                       \\
                                                                     & Discernment   & .113$^{\ast\ast} $ & .128$^{\ast\ast}$                                  \\
                                                                     \midrule
\multirow{3}{*}{\abc
} & Negative      & -0.075                      & -0.021                                                      \\
                                                                     & Positive      & 0.048                       & .11$^{\ast\ast}$                                 \\
                                                                     & Discernment   & 0.088$^\ast $      & .11$^{\ast\ast}$                                  \\
                                                                     \midrule
Total                                                                & Discernment   & .125$^{\ast\ast}$  & .154$^{\ast\ast}$                                  \\ \bottomrule
\end{tabular}
\begin{tablenotes}
{\small 
\item $^{\ast\ast}$: $p < 0.01 $, $^\ast$: $p <0.05$.
}
\end{tablenotes}
\end{threeparttable}

\end{table}
\myparagraph{CRT Test.}
To answer RQ4, we aim to assess the relationships that exists between workers' background and their performance.
In terms of workers' cognitive skills, we measure CRT performance as the percentage of correct answers given by them. Thus, a higher CRT score is associated to higher  analytical thinking ability~\cite{Frederick2005}.
We compare worker performances across the three scales by means of the standardized calculation of the  $z$-score for each worker and each assessment level.
The $z$-score for each statement represents the performance of crowd workers as compared to others.
The lower the $z$-score for false statements, the stronger the ability of the crowd to identify lies and the higher the $z$-score for true statements, the higher the ability to identify accurate information.
``Discernment'' is then calculated by deducting the $z$-score for false statements from the $z$-score for true statements. This represents the ability of the crowd to distinguish the true from the false~\cite{pennycook2019lazy}.
In this analysis we focus on statements with extreme true or false ground truth labels and discard the `in-between' statements as they do not provide additional evidence on the ability of the crowd to distinguish between true/false information.

Table~\ref{tab:Correlation } shows the results.
First, there is a statistically significant (Spearman's rank-order test), moderate positive correlation between Discernment and CRT score on statements from \politifact and \abc ($r_s(598)= 0.128$, $p = 0.002$ and $r_s(598)= 0.11$, $p = 0.007$ respectively).
This shows that  workers who reflect more perform 
better in identifying pants-on-fire statements of US (local) politicians ($r_s(598)= -0.098$, $p = 0.017$), and identifying true statements of AU (not local) politicians ($r_s(598)= 0.11$, $p = 0.007$).
In general, people with strong analytical abilities (as determined by the CRT test) can better recognize true statements from false($r_s(598)= 0.154$, $p<0.0005$).
Besides, the ability to distinguish true from false increases with age ($r_s(598)= 0.125$, $ p=0.002$).
Older workers perform better in recognizing true statements by US politicians ($r_s(598)= 0.127$, $p=0.02$). 
The level of education and income do not have statistically significant correlation with their judgments.

\myparagraph{Political Background.}
As a Shapiro-Wilk test  ~\cite{shapiro1972approximate} confirms that discernment scores are normally distributed ($p > 0.05$) for groups with diverse political views, we can conducted a one-way ANOVA analysis to determine if the ability to distinguish true from false statements was different across groups.\footnote{As a Levene's Test ~\cite{schultz1985levene} showed that the homogeneity of variances was violated ($p = 0.034$) we used the Welch-Satterthwaite correction ~\cite{satterthwaite1946approximate}  to calculate the degrees of freedom and a Games-Howell post-hoc test ~\cite{ruxton2008time}  to show multiple comparisons.}
Discernment score is statistically significantly different between different political views (Welch's $F(4, 176.735) = 3.451$, $p = 0.01$).
A Games-Howell post-hoc test confirms that the increase of discernment score (\num{0.453}, 95\% CI (\num{0.028} to \num{0.879}))  from conservative (\num{-0.208} ± \num{1.293}) to liberal (\num{0.245} ± \num{1.497}) is statistically significant ($p = 0.03$).
Given these results, we can conclude that crowd workers who have liberal views can better differentiate between false and true statements. Furthermore, there is no statistically significant difference (we use a Kruskal-Wallis H tests ~\cite{kruskal1952use} as Shapiro-Wilk test ~\cite{shapiro1972approximate} $p < 0.05$) in discernment scores based on the political party in which crowd workers explicitly identified themselves with ($x^2(3)= 3.548$, $ p= 0.315$).
This shows there is no difference in judgment quality based on their explicit political stance.
However, an analysis of their implicit political views rather than their explicit party identification shows a different result.

From the above results we can see that while the explicit party identification does not have a significant impact on the judgment quality for true and false statements,  their implicit political orientation does. 
The partisan gap on the immigration issue is apparent in the US. According to the survey conducted by Pew Research Center
in January 2019, about 82\% of Republicans and Republican-leaning independents support developing the barrier along the southern border of the United States, while 93\% of Democrats and Democratic leaners oppose it.
Therefore, asking workers' opinion on this matter can be a way to know their implicit political orientation. 

 We use a Kruskal-Wallis H tests ~\cite{kruskal1952use} as Shapiro-Wilk test~\cite{shapiro1972approximate} ($p < 0.05$) in assessor's discernment between immigration policy groups defined based on their answer to the wall question. We observed statistically significant differences on Politifact statements ($x_2 (2) = 10.965, p = 0.004$) and on all statements ($x_2 (2) = 11.966, p = 0.003$).
A post-hoc analysis using Dunn's procedure with a Bonferroni correction ~\cite{dunn1964multiple} revealed statistically significant differences in discernment scores on \politifact statements between agreeing (\num{-0.335}) and disagreeing (\num{0.245}) ($p = 0.007$) on building a wall. 
Similar results are obtained when discernment scores are compared on all statement, there were statistically significant differences in discernment scores between agreeing (\num{-0.377}) and disagreeing (\num{0.173}) ($p = 0.002$) on building a wall along the southern border of the United States. 
These results show how, in our experiment, crowd workers who do not want a wall on the southern US border perform better in distinguishing between true and false statements.
We did not observe significant differences based on their stance on climate issues.

%% file: 5-concl.tex
\section{Discussion and Conclusions}\label{sec:conc}

In this paper, we present an extensive crowdsourcing experiment that aims at studying the overview of crowd assessor identifying misinformation online. The dataset we employed in the research includes statements given by US 
and Australian politicians. The experiment asks US-based crowd workers to perform the fact-checking task by using the customized Internet search engine under our control to find evidence of the validity of the statements. We collect and analyze data on the assessors' political background and cognitive abilities, and control for the politically-consistent statements to be fact-checked, the geographical relevance of the statements, the assessment scale granularity, and the truthfulness level.

Addressing RQ2, in terms of agreement w.r.t. the ground truth, the behavior over all
the three scales is similar, both on \politifact and \abc statements.
Across all the annotation created using crowdsourcing, there is low level of internal agreement among workers, on all the \three, \six, and \onehundred scales. However, addressing RQ1, we found that the grouping of adjacent categories reveals that crowd sourced truthfulness judgments are useful to accurately single out true from false statements.
Addressing RQ3, we found that workers put effort to find a reliable source to justify their judgments, and tend to choose a source found in the first search engine result page, but not necessarily the first search result. 
Finally, concerning RQ4, we found that assessors' background affects in objectively identify online misinformation.

Future work includes a thorough study of the perceived distance between the truthfulness scales, which would inform more sophisticated ways for aggregating and merging crowdsourced judgments. The resource we created can also be used to better understand --via user studies-- how the agreement obtained in crowdsourced judgments can assist experts in better identifying statements that need more attention. We envisage that a collaborative process between automatic credibility systems, crowd workers, and expert fact checkers would provide a scalable and decentralized hybrid mechanism to cope with the increasing volume of online misinformation.

%% file: zzz-main-sigconf.bbl

\begin{thebibliography}{57}


\ifx \showCODEN    \undefined \def \showCODEN     #1{\unskip}     \fi
\ifx \showDOI      \undefined \def \showDOI       #1{#1}\fi
\ifx \showISBNx    \undefined \def \showISBNx     #1{\unskip}     \fi
\ifx \showISBNxiii \undefined \def \showISBNxiii  #1{\unskip}     \fi
\ifx \showISSN     \undefined \def \showISSN      #1{\unskip}     \fi
\ifx \showLCCN     \undefined \def \showLCCN      #1{\unskip}     \fi
\ifx \shownote     \undefined \def \shownote      #1{#1}          \fi
\ifx \showarticletitle \undefined \def \showarticletitle #1{#1}   \fi
\ifx \showURL      \undefined \def \showURL       {\relax}        \fi
\providecommand\bibfield[2]{#2}
\providecommand\bibinfo[2]{#2}
\providecommand\natexlab[1]{#1}
\providecommand\showeprint[2][]{arXiv:#2}

\bibitem[\protect\citeauthoryear{Agresti}{Agresti}{2010}]%
        {agresti2010analysis}
\bibfield{author}{\bibinfo{person}{Alan Agresti}.}
  \bibinfo{year}{2010}\natexlab{}.
\newblock \bibinfo{booktitle}{\emph{Analysis of Ordinal Categorical Data}
  (\bibinfo{edition}{2nd} ed.)}.
\newblock \bibinfo{publisher}{Wiley}.
\newblock



\bibitem[\protect\citeauthoryear{Aletras, Baldwin, Lau, and Stevenson}{Aletras
  et~al\mbox{.}}{2017}]%
        {aletras2017evaluating}
\bibfield{author}{\bibinfo{person}{Nikolaos Aletras}, \bibinfo{person}{Timothy
  Baldwin}, \bibinfo{person}{Jey~Han Lau}, {and} \bibinfo{person}{Mark
  Stevenson}.} \bibinfo{year}{2017}\natexlab{}.
\newblock \showarticletitle{Evaluating Topic Representations for Exploring
  Document Collections}.
\newblock \bibinfo{journal}{\emph{Journal of the Association for Information
  Science and Technology}} \bibinfo{volume}{68}, \bibinfo{number}{1}
  (\bibinfo{year}{2017}), \bibinfo{pages}{154--167}.
\newblock


\bibitem[\protect\citeauthoryear{Amig{\'o}, Carrillo~de Albornoz, Chugur,
  Corujo, Gonzalo, Mart{\'i}n, Meij, de~Rijke, and Spina}{Amig{\'o}
  et~al\mbox{.}}{2013a}]%
        {replab2013}
\bibfield{author}{\bibinfo{person}{Enrique Amig{\'o}}, \bibinfo{person}{Jorge
  Carrillo~de Albornoz}, \bibinfo{person}{Irina Chugur},
  \bibinfo{person}{Adolfo Corujo}, \bibinfo{person}{Julio Gonzalo},
  \bibinfo{person}{Tamara Mart{\'i}n}, \bibinfo{person}{Edgar Meij},
  \bibinfo{person}{Maarten de Rijke}, {and} \bibinfo{person}{Damiano Spina}.}
  \bibinfo{year}{2013}\natexlab{a}.
\newblock \showarticletitle{Overview of {RepLab} 2013: Evaluating Online
  Reputation Monitoring Systems}. In \bibinfo{booktitle}{\emph{Proceedings of
  CLEF}}. \bibinfo{pages}{333--352}.
\newblock


\bibitem[\protect\citeauthoryear{Amig{\'o}, Gonzalo, and Verdejo}{Amig{\'o}
  et~al\mbox{.}}{2013b}]%
        {amigo2013general}
\bibfield{author}{\bibinfo{person}{Enrique Amig{\'o}}, \bibinfo{person}{Julio
  Gonzalo}, {and} \bibinfo{person}{Felisa Verdejo}.}
  \bibinfo{year}{2013}\natexlab{b}.
\newblock \showarticletitle{A General Evaluation Measure for Document
  Organization Tasks}. In \bibinfo{booktitle}{\emph{Proceedings of SIGIR}}.
  \bibinfo{pages}{643--652}.
\newblock


\bibitem[\protect\citeauthoryear{Atanasova, Nakov, M\`{a}rquez,
  Barr\'{o}n-Cede\~{n}o, Karadzhov, Mihaylova, Mohtarami, and Glass}{Atanasova
  et~al\mbox{.}}{2019}]%
        {atanasova2019automatic}
\bibfield{author}{\bibinfo{person}{Pepa Atanasova}, \bibinfo{person}{Preslav
  Nakov}, \bibinfo{person}{Llu\'{\i}s M\`{a}rquez}, \bibinfo{person}{Alberto
  Barr\'{o}n-Cede\~{n}o}, \bibinfo{person}{Georgi Karadzhov},
  \bibinfo{person}{Tsvetomila Mihaylova}, \bibinfo{person}{Mitra Mohtarami},
  {and} \bibinfo{person}{James Glass}.} \bibinfo{year}{2019}\natexlab{}.
\newblock \showarticletitle{Automatic Fact-Checking Using Context and Discourse
  Information}.
\newblock \bibinfo{journal}{\emph{J. Data and Information Quality}}
  \bibinfo{volume}{11}, \bibinfo{number}{3}, Article \bibinfo{articleno}{12}
  (\bibinfo{year}{2019}), \bibinfo{numpages}{27}~pages.
\newblock


\bibitem[\protect\citeauthoryear{Checco, Roitero, Maddalena, Mizzaro, and
  Demartini}{Checco et~al\mbox{.}}{2017}]%
        {checco2017let}
\bibfield{author}{\bibinfo{person}{Alessandro Checco}, \bibinfo{person}{Kevin
  Roitero}, \bibinfo{person}{Eddy Maddalena}, \bibinfo{person}{Stefano
  Mizzaro}, {and} \bibinfo{person}{Gianluca Demartini}.}
  \bibinfo{year}{2017}\natexlab{}.
\newblock \showarticletitle{Let's Agree to Disagree: Fixing Agreement Measures
  for Crowdsourcing}. In \bibinfo{booktitle}{\emph{Proceedings of HCOMP}}.
  \bibinfo{pages}{11--20}.
\newblock


\bibitem[\protect\citeauthoryear{Conroy, Rubin, and Chen}{Conroy
  et~al\mbox{.}}{2015}]%
        {conroy2015automatic}
\bibfield{author}{\bibinfo{person}{Niall~J Conroy}, \bibinfo{person}{Victoria~L
  Rubin}, {and} \bibinfo{person}{Yimin Chen}.} \bibinfo{year}{2015}\natexlab{}.
\newblock \showarticletitle{Automatic Deception Detection: Methods for Finding
  Fake News}.
\newblock \bibinfo{journal}{\emph{Proceedings of the Association for
  Information Science and Technology}} \bibinfo{volume}{52},
  \bibinfo{number}{1} (\bibinfo{year}{2015}), \bibinfo{pages}{1--4}.
\newblock


\bibitem[\protect\citeauthoryear{Craswell, Zoeter, Taylor, and Ramsey}{Craswell
  et~al\mbox{.}}{2008}]%
        {craswell2008experimental}
\bibfield{author}{\bibinfo{person}{Nick Craswell}, \bibinfo{person}{Onno
  Zoeter}, \bibinfo{person}{Michael Taylor}, {and} \bibinfo{person}{Bill
  Ramsey}.} \bibinfo{year}{2008}\natexlab{}.
\newblock \showarticletitle{An Experimental Comparison of Click Position-Bias
  Models}. In \bibinfo{booktitle}{\emph{Proceedings of WSDM}}.
  \bibinfo{pages}{87--94}.
\newblock


\bibitem[\protect\citeauthoryear{Cuzzocrea, Bonchi, and Gunopulos}{Cuzzocrea
  et~al\mbox{.}}{2019}]%
        {DBLP:conf/cikm/2018w}
\bibfield{editor}{\bibinfo{person}{Alfredo Cuzzocrea},
  \bibinfo{person}{Francesco Bonchi}, {and} \bibinfo{person}{Dimitrios
  Gunopulos}} (Eds.). \bibinfo{year}{2019}\natexlab{}.
\newblock \bibinfo{booktitle}{\emph{Proceedings of the {CIKM} 2018 Workshops
  co-located with 27th {ACM CIKM}}}. \bibinfo{series}{{CEUR} Workshop
  Proceedings}, Vol.~\bibinfo{volume}{2482}. \bibinfo{publisher}{CEUR-WS.org}.
\newblock


\bibitem[\protect\citeauthoryear{Dunn}{Dunn}{1964}]%
        {dunn1964multiple}
\bibfield{author}{\bibinfo{person}{Olive~Jean Dunn}.}
  \bibinfo{year}{1964}\natexlab{}.
\newblock \showarticletitle{Multiple Comparisons Using Rank Sums}.
\newblock \bibinfo{journal}{\emph{Technometrics}} \bibinfo{volume}{6},
  \bibinfo{number}{3} (\bibinfo{year}{1964}), \bibinfo{pages}{241--252}.
\newblock


\bibitem[\protect\citeauthoryear{Eickhoff}{Eickhoff}{2018}]%
        {eickhoff2018cognitive}
\bibfield{author}{\bibinfo{person}{Carsten Eickhoff}.}
  \bibinfo{year}{2018}\natexlab{}.
\newblock \showarticletitle{Cognitive Biases in Crowdsourcing}. In
  \bibinfo{booktitle}{\emph{Proceedings of WSDM}}. \bibinfo{pages}{162--170}.
\newblock


\bibitem[\protect\citeauthoryear{Elsayed, Nakov, Barr{\'o}n-Cede{\~n}o,
  Hasanain, Suwaileh, Da~San~Martino, and Atanasova}{Elsayed
  et~al\mbox{.}}{2019}]%
        {elsayed2019overview}
\bibfield{author}{\bibinfo{person}{Tamer Elsayed}, \bibinfo{person}{Preslav
  Nakov}, \bibinfo{person}{Alberto Barr{\'o}n-Cede{\~n}o},
  \bibinfo{person}{Maram Hasanain}, \bibinfo{person}{Reem Suwaileh},
  \bibinfo{person}{Giovanni Da~San~Martino}, {and} \bibinfo{person}{Pepa
  Atanasova}.} \bibinfo{year}{2019}\natexlab{}.
\newblock \showarticletitle{Overview of the {CLEF-2019 CheckThat! Lab}:
  Automatic Identification and Verification of Claims}. In
  \bibinfo{booktitle}{\emph{Proceedings of CLEF}}. \bibinfo{pages}{301--321}.
\newblock


\bibitem[\protect\citeauthoryear{Frederick}{Frederick}{2005}]%
        {Frederick2005}
\bibfield{author}{\bibinfo{person}{Shane Frederick}.}
  \bibinfo{year}{2005}\natexlab{}.
\newblock \showarticletitle{Cognitive Reflection and Decision Making}.
\newblock \bibinfo{journal}{\emph{Journal of Economic Perspectives}}
  \bibinfo{volume}{19}, \bibinfo{number}{4} (\bibinfo{date}{December}
  \bibinfo{year}{2005}), \bibinfo{pages}{25--42}.
\newblock


\bibitem[\protect\citeauthoryear{Fuhr}{Fuhr}{2018}]%
        {Fuhr:2018}
\bibfield{author}{\bibinfo{person}{Norbert Fuhr}.}
  \bibinfo{year}{2018}\natexlab{}.
\newblock \showarticletitle{Some Common Mistakes In {IR} Evaluation, And How
  They Can Be Avoided}.
\newblock \bibinfo{journal}{\emph{SIGIR Forum}} \bibinfo{volume}{51},
  \bibinfo{number}{3} (\bibinfo{year}{2018}), \bibinfo{pages}{32--41}.
\newblock


\bibitem[\protect\citeauthoryear{Gencheva, Nakov, M{\`a}rquez,
  Barr{\'o}n-Cede{\~n}o, and Koychev}{Gencheva et~al\mbox{.}}{2017}]%
        {gencheva2017context}
\bibfield{author}{\bibinfo{person}{Pepa Gencheva}, \bibinfo{person}{Preslav
  Nakov}, \bibinfo{person}{Llu{\'\i}s M{\`a}rquez}, \bibinfo{person}{Alberto
  Barr{\'o}n-Cede{\~n}o}, {and} \bibinfo{person}{Ivan Koychev}.}
  \bibinfo{year}{2017}\natexlab{}.
\newblock \showarticletitle{A Context-Aware Approach for Detecting
  Worth-Checking Claims in Political Debates}. In
  \bibinfo{booktitle}{\emph{Proceedings of the International Conference Recent
  Advances in Natural Language Processing, (RANLP'17)}}.
  \bibinfo{pages}{267--276}.
\newblock


\bibitem[\protect\citeauthoryear{Ghosh, Li, Veale, Rosso, Shutova, Barnden, and
  Reyes}{Ghosh et~al\mbox{.}}{2015}]%
        {Ghosh-15}
\bibfield{author}{\bibinfo{person}{Aniruddha Ghosh}, \bibinfo{person}{Guofu
  Li}, \bibinfo{person}{Tony Veale}, \bibinfo{person}{Paolo Rosso},
  \bibinfo{person}{Ekaterina Shutova}, \bibinfo{person}{John Barnden}, {and}
  \bibinfo{person}{Antonio Reyes}.} \bibinfo{year}{2015}\natexlab{}.
\newblock \showarticletitle{{S}em{E}val-2015 Task 11: Sentiment Analysis of
  Figurative Language in Twitter}. In \bibinfo{booktitle}{\emph{Proceedings of
  SemEval}}. \bibinfo{pages}{470--478}.
\newblock


\bibitem[\protect\citeauthoryear{Graham, Baldwin, and Mathur}{Graham
  et~al\mbox{.}}{2015}]%
        {graham2015accurate}
\bibfield{author}{\bibinfo{person}{Yvette Graham}, \bibinfo{person}{Timothy
  Baldwin}, {and} \bibinfo{person}{Nitika Mathur}.}
  \bibinfo{year}{2015}\natexlab{}.
\newblock \showarticletitle{Accurate Evaluation of Segment-level Machine
  Translation Metrics}. In \bibinfo{booktitle}{\emph{Proceedings of
  NAACL-HLT}}. \bibinfo{pages}{1183--1191}.
\newblock


\bibitem[\protect\citeauthoryear{Han, Roitero, Gadiraju, Sarasua, Checco,
  Maddalena, and Demartini}{Han et~al\mbox{.}}{2019a}]%
        {8873609}
\bibfield{author}{\bibinfo{person}{Lei Han}, \bibinfo{person}{Kevin Roitero},
  \bibinfo{person}{Ujwal Gadiraju}, \bibinfo{person}{Cristina Sarasua},
  \bibinfo{person}{Alessandro Checco}, \bibinfo{person}{Eddy Maddalena}, {and}
  \bibinfo{person}{Gianluca Demartini}.} \bibinfo{year}{2019}\natexlab{a}.
\newblock \showarticletitle{The Impact of Task Abandonment in Crowdsourcing}.
\newblock \bibinfo{journal}{\emph{IEEE Transactions on Knowledge and Data
  Engineering}} (\bibinfo{year}{2019}), \bibinfo{pages}{1--1}.
\newblock


\bibitem[\protect\citeauthoryear{Han, Roitero, Maddalena, Mizzaro, and
  Demartini}{Han et~al\mbox{.}}{2019b}]%
        {scale}
\bibfield{author}{\bibinfo{person}{Lei Han}, \bibinfo{person}{Kevin Roitero},
  \bibinfo{person}{Eddy Maddalena}, \bibinfo{person}{Stefano Mizzaro}, {and}
  \bibinfo{person}{Gianluca Demartini}.} \bibinfo{year}{2019}\natexlab{b}.
\newblock \showarticletitle{On Transforming Relevance Scales}. In
  \bibinfo{booktitle}{\emph{Proceedings of CIKM}}. \bibinfo{pages}{39--48}.
\newblock


\bibitem[\protect\citeauthoryear{Hube, Fetahu, and Gadiraju}{Hube
  et~al\mbox{.}}{2019}]%
        {hube2019understanding}
\bibfield{author}{\bibinfo{person}{Christoph Hube}, \bibinfo{person}{Besnik
  Fetahu}, {and} \bibinfo{person}{Ujwal Gadiraju}.}
  \bibinfo{year}{2019}\natexlab{}.
\newblock \showarticletitle{Understanding and Mitigating Worker Biases in the
  Crowdsourced Collection of Subjective Judgments}. In
  \bibinfo{booktitle}{\emph{Proceedings of CHI}}. 12.
\newblock


\bibitem[\protect\citeauthoryear{Joachims, Granka, Pan, Hembrooke, and
  Gay}{Joachims et~al\mbox{.}}{2005}]%
        {joachims2017accurately}
\bibfield{author}{\bibinfo{person}{Thorsten Joachims}, \bibinfo{person}{Laura
  Granka}, \bibinfo{person}{Bing Pan}, \bibinfo{person}{Helene Hembrooke},
  {and} \bibinfo{person}{Geri Gay}.} \bibinfo{year}{2005}\natexlab{}.
\newblock \showarticletitle{Accurately Interpreting Clickthrough Data as
  Implicit Feedback}. In \bibinfo{booktitle}{\emph{Proceedings of SIGIR}}.
  \bibinfo{pages}{154--161}.
\newblock


\bibitem[\protect\citeauthoryear{Kando}{Kando}{2008}]%
        {Kando-08}
\bibfield{editor}{\bibinfo{person}{Noriko Kando}} (Ed.).
  \bibinfo{year}{2008}\natexlab{}.
\newblock \bibinfo{booktitle}{\emph{Proceedings of the 7th {NTCIR} Workshop
  Meeting on Evaluation of Information Access Technologies: Information
  Retrieval, Question Answering and Clross-Lingual Information Access,
  NTCIR-7}}. \bibinfo{publisher}{{NII}}.
\newblock


\bibitem[\protect\citeauthoryear{Kelly and Azzopardi}{Kelly and
  Azzopardi}{2015}]%
        {kelly2015how}
\bibfield{author}{\bibinfo{person}{Diane Kelly} {and} \bibinfo{person}{Leif
  Azzopardi}.} \bibinfo{year}{2015}\natexlab{}.
\newblock \showarticletitle{How Many Results per Page? A Study of SERP Size,
  Search Behavior and User Experience}. In
  \bibinfo{booktitle}{\emph{Proceedings of SIGIR}}. \bibinfo{pages}{183--192}.
\newblock


\bibitem[\protect\citeauthoryear{Kim, Kim, and Oh}{Kim et~al\mbox{.}}{2019}]%
        {kim2019homogeneity}
\bibfield{author}{\bibinfo{person}{Jooyeon Kim}, \bibinfo{person}{Dongkwan
  Kim}, {and} \bibinfo{person}{Alice Oh}.} \bibinfo{year}{2019}\natexlab{}.
\newblock \showarticletitle{Homogeneity-Based Transmissive Process to Model
  True and False News in Social Networks}. \bibinfo{pages}{348--356}.
\newblock


\bibitem[\protect\citeauthoryear{Kriplean, Bonnar, Borning, Kinney, and
  Gill}{Kriplean et~al\mbox{.}}{2014}]%
        {Kriplean:2014:IOF:2531602.2531677}
\bibfield{author}{\bibinfo{person}{Travis Kriplean}, \bibinfo{person}{Caitlin
  Bonnar}, \bibinfo{person}{Alan Borning}, \bibinfo{person}{Bo Kinney}, {and}
  \bibinfo{person}{Brian Gill}.} \bibinfo{year}{2014}\natexlab{}.
\newblock \showarticletitle{Integrating On-demand Fact-checking with Public
  Dialogue}. In \bibinfo{booktitle}{\emph{Proceedings of CSCW}}.
  \bibinfo{pages}{1188--1199}.
\newblock


\bibitem[\protect\citeauthoryear{Krippendorff}{Krippendorff}{2011}]%
        {krippendorff2011computing}
\bibfield{author}{\bibinfo{person}{Klaus Krippendorff}.}
  \bibinfo{year}{2011}\natexlab{}.
\newblock \showarticletitle{Computing Krippendorff's Alpha-Reliability}.
\newblock  (\bibinfo{year}{2011}).
\newblock


\bibitem[\protect\citeauthoryear{Kruskal and Wallis}{Kruskal and
  Wallis}{1952}]%
        {kruskal1952use}
\bibfield{author}{\bibinfo{person}{William~H Kruskal} {and}
  \bibinfo{person}{W~Allen Wallis}.} \bibinfo{year}{1952}\natexlab{}.
\newblock \showarticletitle{Use of Ranks in One-criterion Variance Analysis}.
\newblock \bibinfo{journal}{\emph{Journal of the American statistical
  Association}} \bibinfo{volume}{47}, \bibinfo{number}{260}
  (\bibinfo{year}{1952}).
\newblock


\bibitem[\protect\citeauthoryear{{La Barbera}, Roitero, Spina, Mizzaro, and
  Demartini}{{La Barbera} et~al\mbox{.}}{2020}]%
        {labarbera2020crowdsourcing}
\bibfield{author}{\bibinfo{person}{David {La Barbera}}, \bibinfo{person}{Kevin
  Roitero}, \bibinfo{person}{Damiano Spina}, \bibinfo{person}{Stefano Mizzaro},
  {and} \bibinfo{person}{Gianluca Demartini}.} \bibinfo{year}{2020}\natexlab{}.
\newblock \showarticletitle{{Crowdsourcing Truthfulness: The Impact of Judgment
  Scale and Assessor Bias}}. In \bibinfo{booktitle}{\emph{Proceedings of
  ECIR}}. \bibinfo{pages}{207--214}.
\newblock


\bibitem[\protect\citeauthoryear{Maddalena, Ceolin, and Mizzaro}{Maddalena
  et~al\mbox{.}}{2018}]%
        {INRA:2018}
\bibfield{author}{\bibinfo{person}{Eddy Maddalena}, \bibinfo{person}{Davide
  Ceolin}, {and} \bibinfo{person}{Stefano Mizzaro}.}
  \bibinfo{year}{2018}\natexlab{}.
\newblock \showarticletitle{Multidimensional News Quality: {A} Comparison of
  Crowdsourcing and Nichesourcing}, See \citeN{DBLP:conf/cikm/2018w}.
\newblock


\bibitem[\protect\citeauthoryear{Maddalena, Mizzaro, Scholer, and
  Turpin}{Maddalena et~al\mbox{.}}{2017a}]%
        {Maddalena:2017:CRM:3026478.3002172}
\bibfield{author}{\bibinfo{person}{Eddy Maddalena}, \bibinfo{person}{Stefano
  Mizzaro}, \bibinfo{person}{Falk Scholer}, {and} \bibinfo{person}{Andrew
  Turpin}.} \bibinfo{year}{2017}\natexlab{a}.
\newblock \showarticletitle{On Crowdsourcing Relevance Magnitudes for
  Information Retrieval Evaluation}.
\newblock \bibinfo{journal}{\emph{ACM Transactions on Information Systems}}
  \bibinfo{volume}{35}, \bibinfo{number}{3}, Article \bibinfo{articleno}{19}
  (\bibinfo{year}{2017}), \bibinfo{numpages}{32}~pages.
\newblock


\bibitem[\protect\citeauthoryear{Maddalena, Roitero, Demartini, and
  Mizzaro}{Maddalena et~al\mbox{.}}{2017b}]%
        {10.1145/3121050.3121060}
\bibfield{author}{\bibinfo{person}{Eddy Maddalena}, \bibinfo{person}{Kevin
  Roitero}, \bibinfo{person}{Gianluca Demartini}, {and}
  \bibinfo{person}{Stefano Mizzaro}.} \bibinfo{year}{2017}\natexlab{b}.
\newblock \showarticletitle{Considering Assessor Agreement in IR Evaluation}.
  In \bibinfo{booktitle}{\emph{Proceedings of ICTIR}}.
  \bibinfo{pages}{75--82}.
\newblock


\bibitem[\protect\citeauthoryear{Mathur, Baldwin, and Cohn}{Mathur
  et~al\mbox{.}}{2017}]%
        {mathur2017sequence}
\bibfield{author}{\bibinfo{person}{Nitika Mathur}, \bibinfo{person}{Timothy
  Baldwin}, {and} \bibinfo{person}{Trevor Cohn}.}
  \bibinfo{year}{2017}\natexlab{}.
\newblock \showarticletitle{Sequence Effects in Crowdsourced Annotations}. In
  \bibinfo{booktitle}{\emph{Proceedings of EMNLP}}.
  \bibinfo{pages}{2860--2865}.
\newblock


\bibitem[\protect\citeauthoryear{Mihaylova, Karadjov, Atanasova, Baly,
  Mohtarami, and Nakov}{Mihaylova et~al\mbox{.}}{2019}]%
        {mihaylova2019semeval}
\bibfield{author}{\bibinfo{person}{Tsvetomila Mihaylova},
  \bibinfo{person}{Georgi Karadjov}, \bibinfo{person}{Pepa Atanasova},
  \bibinfo{person}{Ramy Baly}, \bibinfo{person}{Mitra Mohtarami}, {and}
  \bibinfo{person}{Preslav Nakov}.} \bibinfo{year}{2019}\natexlab{}.
\newblock \showarticletitle{SemEval-2019 Task 8: Fact Checking in Community
  Question Answering Forums}. In \bibinfo{booktitle}{\emph{Proceedings of
  SemEval}}. \bibinfo{pages}{860--869}.
\newblock


\bibitem[\protect\citeauthoryear{Nakov, Barr\'{o}n-Cede\~{n}o, Elsayed,
  Suwaileh, M\`{a}rquez, Zaghouani, Atanasova, Kyuchukov, and
  Da~San~Martino}{Nakov et~al\mbox{.}}{2018}]%
        {clef2018checkthat}
\bibfield{author}{\bibinfo{person}{Preslav Nakov}, \bibinfo{person}{Alberto
  Barr\'{o}n-Cede\~{n}o}, \bibinfo{person}{Tamer Elsayed},
  \bibinfo{person}{Reem Suwaileh}, \bibinfo{person}{Llu\'{i}s M\`{a}rquez},
  \bibinfo{person}{Wajdi Zaghouani}, \bibinfo{person}{Pepa Atanasova},
  \bibinfo{person}{Spas Kyuchukov}, {and} \bibinfo{person}{Giovanni
  Da~San~Martino}.} \bibinfo{year}{2018}\natexlab{}.
\newblock \showarticletitle{Overview of the {CLEF-2018 CheckThat! Lab} on
  Automatic Identification and Verification of Political Claims}. In
  \bibinfo{booktitle}{\emph{Proceedings of CLEF}}. \bibinfo{pages}{372--387}.
\newblock


\bibitem[\protect\citeauthoryear{Otterbacher, Bates, and Clough}{Otterbacher
  et~al\mbox{.}}{2017}]%
        {otterbacher2017competent}
\bibfield{author}{\bibinfo{person}{Jahna Otterbacher}, \bibinfo{person}{Jo
  Bates}, {and} \bibinfo{person}{Paul Clough}.}
  \bibinfo{year}{2017}\natexlab{}.
\newblock \showarticletitle{Competent Men and Warm Women: Gender Stereotypes
  and Backlash in Image Search Results}. In
  \bibinfo{booktitle}{\emph{Proceedings of CHI}}. \bibinfo{pages}{6620--6631}.
\newblock


\bibitem[\protect\citeauthoryear{Pennycook and Rand}{Pennycook and
  Rand}{2018}]%
        {pennycook2018falls}
\bibfield{author}{\bibinfo{person}{Gordon Pennycook} {and}
  \bibinfo{person}{David~G Rand}.} \bibinfo{year}{2018}\natexlab{}.
\newblock \showarticletitle{Who Falls for Fake News? The Roles of Bullshit
  Receptivity, Overclaiming, Familiarity, and Analytic Thinking}.
\newblock \bibinfo{journal}{\emph{Journal of Personality}}
  (\bibinfo{year}{2018}), 63.
\newblock


\bibitem[\protect\citeauthoryear{Pennycook and Rand}{Pennycook and
  Rand}{2019a}]%
        {pennycook2019fighting}
\bibfield{author}{\bibinfo{person}{Gordon Pennycook} {and}
  \bibinfo{person}{David~G Rand}.} \bibinfo{year}{2019}\natexlab{a}.
\newblock \showarticletitle{Fighting Misinformation on Social Media Using
  Crowdsourced Judgments of News Source Quality}.
\newblock \bibinfo{journal}{\emph{Proceedings of the National Academy of
  Sciences}} \bibinfo{volume}{116}, \bibinfo{number}{7} (\bibinfo{year}{2019}),
  \bibinfo{pages}{2521--2526}.
\newblock


\bibitem[\protect\citeauthoryear{Pennycook and Rand}{Pennycook and
  Rand}{2019b}]%
        {pennycook2019lazy}
\bibfield{author}{\bibinfo{person}{Gordon Pennycook} {and}
  \bibinfo{person}{David~G Rand}.} \bibinfo{year}{2019}\natexlab{b}.
\newblock \showarticletitle{Lazy, Not Biased: Susceptibility to Partisan Fake
  News Is Better Explained by Lack of Reasoning than by Motivated Reasoning}.
\newblock \bibinfo{journal}{\emph{Cognition}}  \bibinfo{volume}{188}
  (\bibinfo{year}{2019}), \bibinfo{pages}{39--50}.
\newblock


\bibitem[\protect\citeauthoryear{Popat}{Popat}{2019}]%
        {Popat_2019}
\bibfield{author}{\bibinfo{person}{Kashyap~Kiritbhai Popat}.}
  \bibinfo{year}{2019}\natexlab{}.
\newblock \emph{\bibinfo{title}{Credibility Analysis of Textual Claims with
  Explainable Evidence}}.
\newblock \bibinfo{thesistype}{Ph.D. Dissertation}. \bibinfo{school}{Saarland
  University}.
\newblock


\bibitem[\protect\citeauthoryear{Roitero, Demartini, Mizzaro, and
  Spina}{Roitero et~al\mbox{.}}{2018a}]%
        {RSDM:2018}
\bibfield{author}{\bibinfo{person}{Kevin Roitero}, \bibinfo{person}{Gianluca
  Demartini}, \bibinfo{person}{Stefano Mizzaro}, {and} \bibinfo{person}{Damiano
  Spina}.} \bibinfo{year}{2018}\natexlab{a}.
\newblock \showarticletitle{{How Many Truth Levels? Six? One Hundred? Even
  More? Validating Truthfulness of Statements via Crowdsourcing}}, See
  \citeN{DBLP:conf/cikm/2018w}.
\newblock


\bibitem[\protect\citeauthoryear{Roitero, Maddalena, Demartini, and
  Mizzaro}{Roitero et~al\mbox{.}}{2018b}]%
        {Roitero:2018:FRS:3209978.3210052}
\bibfield{author}{\bibinfo{person}{Kevin Roitero}, \bibinfo{person}{Eddy
  Maddalena}, \bibinfo{person}{Gianluca Demartini}, {and}
  \bibinfo{person}{Stefano Mizzaro}.} \bibinfo{year}{2018}\natexlab{b}.
\newblock \showarticletitle{On Fine-Grained Relevance Scales}. In
  \bibinfo{booktitle}{\emph{Proceedings of SIGIR}}. \bibinfo{pages}{675--684}.
\newblock


\bibitem[\protect\citeauthoryear{Ruchansky, Seo, and Liu}{Ruchansky
  et~al\mbox{.}}{2017}]%
        {ruchansky2017csi}
\bibfield{author}{\bibinfo{person}{Natali Ruchansky}, \bibinfo{person}{Sungyong
  Seo}, {and} \bibinfo{person}{Yan Liu}.} \bibinfo{year}{2017}\natexlab{}.
\newblock \showarticletitle{CSI: A Hybrid Deep Model for Fake News Detection}.
  In \bibinfo{booktitle}{\emph{Proceedings of CIKM}}.
  \bibinfo{pages}{797--806}.
\newblock


\bibitem[\protect\citeauthoryear{Ruxton and Beauchamp}{Ruxton and
  Beauchamp}{2008}]%
        {ruxton2008time}
\bibfield{author}{\bibinfo{person}{Graeme~D Ruxton} {and} \bibinfo{person}{Guy
  Beauchamp}.} \bibinfo{year}{2008}\natexlab{}.
\newblock \showarticletitle{Time for Some A Priori Thinking About Post Hoc
  Testing}.
\newblock \bibinfo{journal}{\emph{Behavioral ecology}} \bibinfo{volume}{19},
  \bibinfo{number}{3} (\bibinfo{year}{2008}), \bibinfo{pages}{690--693}.
\newblock


\bibitem[\protect\citeauthoryear{Satterthwaite}{Satterthwaite}{1946}]%
        {satterthwaite1946approximate}
\bibfield{author}{\bibinfo{person}{Franklin~E Satterthwaite}.}
  \bibinfo{year}{1946}\natexlab{}.
\newblock \showarticletitle{An Approximate Distribution of Estimates of
  Variance Components}.
\newblock \bibinfo{journal}{\emph{Biometrics bulletin}} \bibinfo{volume}{2},
  \bibinfo{number}{6} (\bibinfo{year}{1946}), \bibinfo{pages}{110--114}.
\newblock


\bibitem[\protect\citeauthoryear{Schultz}{Schultz}{1985}]%
        {schultz1985levene}
\bibfield{author}{\bibinfo{person}{Brian~B Schultz}.}
  \bibinfo{year}{1985}\natexlab{}.
\newblock \showarticletitle{Levene's Test for Relative Variation}.
\newblock \bibinfo{journal}{\emph{Systematic Zoology}} \bibinfo{volume}{34},
  \bibinfo{number}{4} (\bibinfo{year}{1985}), \bibinfo{pages}{449--456}.
\newblock


\bibitem[\protect\citeauthoryear{Shapiro and Francia}{Shapiro and
  Francia}{1972}]%
        {shapiro1972approximate}
\bibfield{author}{\bibinfo{person}{Samuel~S Shapiro} {and} \bibinfo{person}{RS
  Francia}.} \bibinfo{year}{1972}\natexlab{}.
\newblock \showarticletitle{An Approximate Analysis of Variance Test for
  Normality}.
\newblock \bibinfo{journal}{\emph{J. Amer. Statist. Assoc.}}
  \bibinfo{volume}{67}, \bibinfo{number}{337} (\bibinfo{year}{1972}),
  \bibinfo{pages}{215--216}.
\newblock


\bibitem[\protect\citeauthoryear{Singhania, Fernandez, and Rao}{Singhania
  et~al\mbox{.}}{2017}]%
        {singhania20173han}
\bibfield{author}{\bibinfo{person}{Sneha Singhania}, \bibinfo{person}{Nigel
  Fernandez}, {and} \bibinfo{person}{Shrisha Rao}.}
  \bibinfo{year}{2017}\natexlab{}.
\newblock \showarticletitle{3HAN: A Deep Neural Network for Fake News
  Detection}. In \bibinfo{booktitle}{\emph{Proceedings of the International
  Conference on Neural Information Processing (ICONIP'17)}}. Springer,
  \bibinfo{pages}{572--581}.
\newblock


\bibitem[\protect\citeauthoryear{Tchechmedjiev, Fafalios, Boland, Gasquet,
  Zloch, Zapilko, Dietze, and Todorov}{Tchechmedjiev et~al\mbox{.}}{2019}]%
        {tchechmedjiev2019claimskg}
\bibfield{author}{\bibinfo{person}{Andon Tchechmedjiev},
  \bibinfo{person}{Pavlos Fafalios}, \bibinfo{person}{Katarina Boland},
  \bibinfo{person}{Malo Gasquet}, \bibinfo{person}{Matth{\"a}us Zloch},
  \bibinfo{person}{Benjamin Zapilko}, \bibinfo{person}{Stefan Dietze}, {and}
  \bibinfo{person}{Konstantin Todorov}.} \bibinfo{year}{2019}\natexlab{}.
\newblock \showarticletitle{ClaimsKG: A Knowledge Graph of Fact-Checked
  Claims}. In \bibinfo{booktitle}{\emph{Proceedings of the International
  Semantic Web Conference}}. \bibinfo{pages}{309--324}.
\newblock


\bibitem[\protect\citeauthoryear{Vasileva, Atanasova, M{\`a}rquez,
  Barr{\'o}n-Cede{\~n}o, and Nakov}{Vasileva et~al\mbox{.}}{2019}]%
        {vasileva2019takes}
\bibfield{author}{\bibinfo{person}{Slavena Vasileva}, \bibinfo{person}{Pepa
  Atanasova}, \bibinfo{person}{Llu{\'\i}s M{\`a}rquez},
  \bibinfo{person}{Alberto Barr{\'o}n-Cede{\~n}o}, {and}
  \bibinfo{person}{Preslav Nakov}.} \bibinfo{year}{2019}\natexlab{}.
\newblock \showarticletitle{It Takes Nine to Smell a Rat: Neural Multi-Task
  Learning for Check-Worthiness Prediction}. In
  \bibinfo{booktitle}{\emph{Proceedings of Recent Advances in Natural Language
  Processing (RANLP'19)}}. \bibinfo{pages}{1229--1239}.
\newblock


\bibitem[\protect\citeauthoryear{Vlachos and Riedel}{Vlachos and
  Riedel}{2014}]%
        {vlachos2014fact}
\bibfield{author}{\bibinfo{person}{Andreas Vlachos} {and}
  \bibinfo{person}{Sebastian Riedel}.} \bibinfo{year}{2014}\natexlab{}.
\newblock \showarticletitle{Fact Checking: Task Definition and Dataset
  Construction}. In \bibinfo{booktitle}{\emph{Proceedings of the ACL 2014
  Workshop on Language Technologies and Computational Social Science}}.
  \bibinfo{pages}{18--22}.
\newblock


\bibitem[\protect\citeauthoryear{Vo and Lee}{Vo and Lee}{2018}]%
        {vo2018rise}
\bibfield{author}{\bibinfo{person}{Nguyen Vo} {and} \bibinfo{person}{Kyumin
  Lee}.} \bibinfo{year}{2018}\natexlab{}.
\newblock \showarticletitle{The Rise of Guardians: Fact-checking URL
  Recommendation to Combat Fake News}. In \bibinfo{booktitle}{\emph{Proceedings
  of SIGIR}}. \bibinfo{pages}{275--284}.
\newblock


\bibitem[\protect\citeauthoryear{Vo and Lee}{Vo and Lee}{2019}]%
        {vo2019learning}
\bibfield{author}{\bibinfo{person}{Nguyen Vo} {and} \bibinfo{person}{Kyumin
  Lee}.} \bibinfo{year}{2019}\natexlab{}.
\newblock \showarticletitle{Learning from Fact-Checkers: Analysis and
  Generation of Fact-Checking Language}. In
  \bibinfo{booktitle}{\emph{Proceedings of SIGIR}}. \bibinfo{pages}{335--344}.
\newblock


\bibitem[\protect\citeauthoryear{Vosoughi, Roy, and Aral}{Vosoughi
  et~al\mbox{.}}{2018}]%
        {vosoughi2018spread}
\bibfield{author}{\bibinfo{person}{Soroush Vosoughi}, \bibinfo{person}{Deb
  Roy}, {and} \bibinfo{person}{Sinan Aral}.} \bibinfo{year}{2018}\natexlab{}.
\newblock \showarticletitle{The Spread of True and False News Online}.
\newblock \bibinfo{journal}{\emph{Science}} \bibinfo{volume}{359},
  \bibinfo{number}{6380} (\bibinfo{year}{2018}), \bibinfo{pages}{1146--1151}.
\newblock


\bibitem[\protect\citeauthoryear{Wang}{Wang}{2017}]%
        {politifact}
\bibfield{author}{\bibinfo{person}{William~Yang Wang}.}
  \bibinfo{year}{2017}\natexlab{}.
\newblock \showarticletitle{``{Liar, Liar Pants on Fire}'': A New Benchmark
  Dataset for Fake News Detection}. In \bibinfo{booktitle}{\emph{Proceedings of
  ACL}}. \bibinfo{pages}{422--426}.
\newblock


\bibitem[\protect\citeauthoryear{Wang, Ma, Jin, Yuan, Xun, Jha, Su, and
  Gao}{Wang et~al\mbox{.}}{2018}]%
        {wang2018eann}
\bibfield{author}{\bibinfo{person}{Yaqing Wang}, \bibinfo{person}{Fenglong Ma},
  \bibinfo{person}{Zhiwei Jin}, \bibinfo{person}{Ye Yuan},
  \bibinfo{person}{Guangxu Xun}, \bibinfo{person}{Kishlay Jha},
  \bibinfo{person}{Lu Su}, {and} \bibinfo{person}{Jing Gao}.}
  \bibinfo{year}{2018}\natexlab{}.
\newblock \showarticletitle{EANN: Event Adversarial Neural Networks for
  Multi-modal Fake News Detection}. In \bibinfo{booktitle}{\emph{Proceedings of
  KDD}}. \bibinfo{pages}{849--857}.
\newblock


\bibitem[\protect\citeauthoryear{You, Vo, Lee, and LIU}{You
  et~al\mbox{.}}{2019}]%
        {you2019attributed}
\bibfield{author}{\bibinfo{person}{Di You}, \bibinfo{person}{Nguyen Vo},
  \bibinfo{person}{Kyumin Lee}, {and} \bibinfo{person}{Qiang LIU}.}
  \bibinfo{year}{2019}\natexlab{}.
\newblock \showarticletitle{Attributed Multi-Relational Attention Network for
  Fact-Checking URL Recommendation}. In \bibinfo{booktitle}{\emph{Proceedings
  of CIKM}}. \bibinfo{pages}{1471--1480}.
\newblock


\bibitem[\protect\citeauthoryear{Zubiaga and Ji}{Zubiaga and Ji}{2014}]%
        {zubiaga2014tweet}
\bibfield{author}{\bibinfo{person}{Arkaitz Zubiaga} {and} \bibinfo{person}{Heng
  Ji}.} \bibinfo{year}{2014}\natexlab{}.
\newblock \showarticletitle{Tweet, but Verify: Epistemic Study of Information
  Verification on Twitter}.
\newblock \bibinfo{journal}{\emph{Social Network Analysis and Mining}}
  \bibinfo{volume}{4}, \bibinfo{number}{1} (\bibinfo{year}{2014}),
  \bibinfo{pages}{1--12}.
\newblock


\end{thebibliography}
